\documentclass[twocolumn,aps,pra,floatfix,superscriptaddress,secnumarabic]{revtex4-1}
\pdfoutput=1
\usepackage{amsmath}
\usepackage{amsfonts}
\usepackage{amssymb}
\usepackage{graphicx} 
\usepackage{bm}
\usepackage{hyperref}
\usepackage{bookmark}
\hypersetup{colorlinks, allcolors={blue}}
\newcommand{\cw}{8.cm}
\newcommand*{\fref}[1]{fig.~\ref{#1}}
\newcommand*{\Fref}[1]{Figure~\ref{#1}}
\newcommand*{\ffref}[1]{figs.~\ref{#1}}

\newcommand*{\sref}[1]{Sec.~\ref{#1}}
\newcommand*{\Sref}[1]{Section~\ref{#1}}

\newcommand*{\cref}[1]{ref.~\cite{#1}}
\newcommand*{\Cref}[1]{Reference~\cite{#1}}
\newcommand*{\ccref}[1]{refs.~\cite{#1}}

\newcommand{\Frac}[2]{\displaystyle\frac{#1}{#2}}
\begin{document}

\author{Alessandro Cresti}
\affiliation{Univ. Grenoble Alpes, Univ. Savoie Mont Blanc, CNRS, Grenoble INP, IMEP-LAHC, 38000 Grenoble, France}
\author{Jes\'us Carrete}
\affiliation{Institute of Materials Chemistry, TU Wien, A-1060 Vienna, Austria}
\author{Hanako Okuno}
\affiliation{Department of Physics, IriG, Univ. Grenoble Alpes and CEA, F-38000 Grenoble, France}
\author{Tao Wang}
\affiliation{ICAMS, Ruhr-Universit\"{a}t Bochum, D-44780 Bochum, Germany}
\author{Georg K. H. Madsen}
\affiliation{Institute of Materials Chemistry, TU Wien, A-1060 Vienna, Austria}
\author{Natalio Mingo}
\affiliation{Univ. Grenoble Alpes, CEA, LITEN, F-38000 Grenoble, France}
\author{Pascal Pochet}
\affiliation{Department of Physics, IriG, Univ. Grenoble Alpes and CEA, F-38000 Grenoble, France}

\title{Growth, charge and thermal transport of flowered graphene} 

\begin{abstract}
We report on the structural and transport properties of the smallest dislocation loop in graphene, known as a flower defect. 
First, by means of advanced experimental imaging techniques, we deduce how flower defects are formed during recrystallization of chemical vapor deposited graphene. 
We propose that the flower defects arise from a bulge type mechanism in which the flower domains are the grains \textit{left over} by dynamic recrystallisation. 
Next, in order to evaluate the use of such defects as possible building blocks for \mbox{all-graphene} electronics, we combine multiscale modeling tools to investigate the structure and the electron and phonon transport properties of large monolayer graphene samples with a random distribution of flower defects. 
For large enough flower densities, we find that electron transport is strongly suppressed while, surprisingly, hole transport remains almost unaffected. 
These results suggest possible applications of flowered graphene for electron energy filtering. 
For the same defect densities, phonon transport is reduced by orders of magnitude as elastic scattering by defects becomes dominant. 
Heat transport by flexural phonons, key in graphene, is largely suppressed even for very low concentrations.
\\[5mm]
\noindent doi: \href{\doibase 10.1016/j.carbon.2020.01.040}{10.1016/j.carbon.2020.01.040}

\end{abstract}

\maketitle

\section{Introduction} \label{Sec:Introduction}

Since its discovery~\cite{NOV_SCI306}, graphene has sparkled an incredible amount of interest for a large spectrum of potential applications~\cite{FER_NS7} including electronics~\cite{LEM_MRS39}, flexible optoelectronics~\cite{BON_NP4}, spintronics~\cite{HAN_NN9}, metrology~\cite{POI_EPJ172,LEH_AO50} and more exotic valleytronics~\cite{RYC_NP3,GOR_SCI346}. 
Some of these applications, e.g. the definition of a resistance standard~\cite{RIB_NN10} or the realization of radio frequency transistors~\cite{WU_NL12,GUO_NL13} and \mbox{light-emitting} diodes~\cite{WIT_NM14}, are already reality and are pushing the industrial research on graphene. 
Other applications are still remote. 
In particular, the use of graphene for digital electronics is compromised by the absence of band gap, which limits the on/off ratio in \mbox{field-effect} transistors in spite of the high mobility of graphene. 
Proposals to open a band or mobility gap include confinement in nanoribbons~\cite{SON_PRL97}, doping~\cite{BIE_NL9,MAR_ACSN6} and use of biased~\cite{CAS_PRL99,OOS_NM7} bilayer graphene. 
To date, however, none of these solutions has proven to be effective due to the narrowness of the gap or the excessive degradation of the charge mobility. 
Another critical issue for applications is the large scale production of \mbox{high-quality} monolayer graphene. 
In this respect, the growth by chemical vapor deposition (CVD)~\cite{MUN_CVD19} represents a booster for graphene industrialization, and a great opportunity to explore new physics, especially related to topological defects. 
Indeed, polycrystalline domains, grain boundaries, dislocations and line defects are typical imperfections of CVD graphene~\cite{HUA_NAT469,COC_PRB83,BAT_SSR67,COC_PRB85,HER_PCL6,BIR_NJP15,TER_RPP75,BAN_ACSN5,LU_ACSN7}, which can strongly affect its transport properties~\cite{YAZ_NM9,LHE_PRB86,VAN_NL13,LAF_PRB90,Vancso2013} depending on the boundary morphology~\cite{TSE_SCI336}.
However, this limitation can be turned into an opportunity if, properly engineered, such defects~\cite{KUR_NL12,LEH_NC4,LEE_JPCM25,JAN_JACS136,WU_AM25,LU_ACSN7,ROB_NC3} are exploited to induce phenomena, as valley filtering~\cite{CHE_PRB89}, of great fundamental and potentially technological interest. 
The increased chemical reactivity of these extended line defects can also find application in gas~\cite{Souza2018} or ion~\cite{Veliev2018} sensors. 
Thermal management is another critical aspect of nanodevice design, so it comes as no surprise that the unique thermal transport properties of graphene have also attracted great attention~\cite{pop_thermal_2012}. 
Pristine graphene exhibits extraordinarily high thermal conductivity (with room temperature values in the \mbox{3000-5000 Wm$^{-1}$K$^{-1}$} range)~\cite{lindsay_phonon_2014,Fugallo_NL14,Xu_NatComm14}.
In pristine systems, the main phonon scattering mechanism is anharmonicity, manifested in \mbox{three-phonon} processes, and graphene owes its high thermal conductivity to the high density of states of its flexural phonon branch at low energies, together with a \mbox{symmetry-induced} selection rule for three-phonon scattering processes~\cite{lindsay_flexural_2010}. 
Several theoretical studies have also addressed the broader problem of thermal transport in \mbox{defect-laden} graphene~\cite{KOL_EPL100, KHO_CMS79, krasavin_effect_2015, zhao_defect-engineered_2015,Hahn2016,Sevim2018,Nobakht2018} and graphene nanostructures~\cite{TAN_CAR65, YAN_PLA377, zhu_mechanisms_2016}. However, in general those studies use either classical molecular dynamics or simple parametric models, both of which fail to give a detailed insight into the phonon physics underpinning the complex transport behavior in these systems.

In this work, we combine experimental imaging techniques -- \mbox{high-resolution} transmission electron microscopy (HRTEM) -- and advanced modeling tools --- density functional theory (DFT), Green's function techniques and the Boltzmann transport equation --- to investigate the structure and the electron and phonon transport properties of graphene with flower defects.
A flower defect~\cite{HUA_NAT469,COC_PRB83} can be seen as a 30$^\circ$ rotation of a group of seven carbon rings around the central ring (see \fref{fig:series}(e)). 
It represents the smallest possible grain boundary loop in graphene and has been envisioned to be exploited as building block for \mbox{all-graphene} nanoelectronics~\cite{YAN_APL103} and spintronics~\cite{Kang2019}. 
It is thus crucial to estimate the impact of this kind of topological defects on the electron and phonon transport properties of CVD graphene, and to understand the related physics, which could potentially lead to original applications. 

While flower defects do not break the sp$^2$ hybridization of the carbon atoms involved, they do introduce pentagonal and heptagonal rings, which break the sublattice symmetry of graphene. 
Therefore, as demonstrated in the rest of the paper, they are expected to strongly affect the transport properties and introduce an \mbox{electron-hole} asymmetry.
While the literature mainly focuses on isolated defects, here we consider random defect distributions over large samples, thus providing a more realistic characterization of their expected behavior. 

The paper is organized as follows. 
\Sref{Sec:Results_STR} illustrates some HRTEM measurements of CVD graphene~\cite{TYU_CA102}, and provides experimental evidence of the presence of flower defects in our samples as well as a possible mechanism behind their growth in our sample. 
In the same section, we show {\it ab initio} calculations to study their energetic stability. 
Triangular aggregates of flowers are also investigated.
In \sref{Sec:Results}, we investigate the electron (\sref{Sec:Results_ELEC}) and the phonon (\sref{Sec:Results_THER}) transport properties of flowered graphene, and provide an insight into the related physics. 
In particular, we show that a high density of flowers can open a transport gap for electrons, while surprisingly leaving the hole transmission almost unaffected. 
Additionally, a prospective application of our results, which mainly relies on the opening of a transport gap for electrons and on the future possibility of engineering the flower defects by a controlled recrystallization of graphene, is discussed.
Moreover, the thermal conductance turns out to be strongly affected and shows a peculiar plateau for moderate defect concentrations, before being strongly suppressed for higher concentrations.
Finally, \sref{Sec:Conclusions} summarizes our results.
The details of our models and simulations are given in the methodology \sref{Sec:Methods}.

\section{Structural properties and growth mechanism of flower defects} \label{Sec:Results_STR}

 \begin{figure*}[!ht]
   \begin{center}
   \resizebox{16cm}{!}{\includegraphics{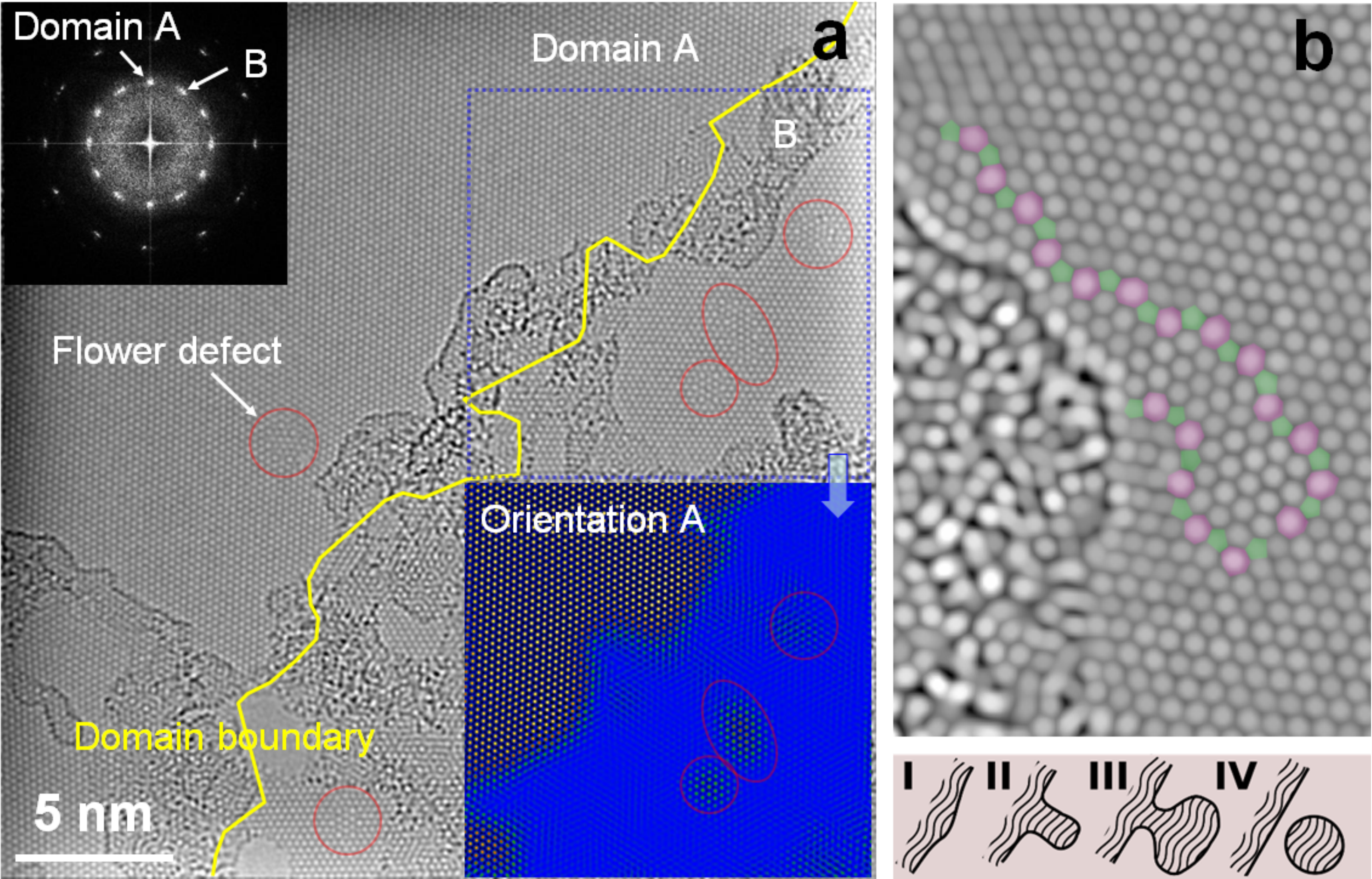}}
  \end{center}
  \caption{
  (a)~HRTEM image of CVD graphene with misoriented domains showing \mbox{flower-related} defects circled in red on both side of the grain boundary (yellow line). 
  Insets:~Fast Fourier Transform (FFT) of the figure in panel (a) and \mbox{dark-field} image of numerically selected domain oriented in A direction shown in FFT. 
  (b)~Identification of a graphene bulge corresponding to stage II or III in the sketched bulge nucleation mechanism. Pentagons and heptagons are highlighted in green and magenta respectively.
  }
  \label{fig:flower} 
 \end{figure*}

\begin{figure*}[!t]
  \begin{center}
  \resizebox{15cm}{!}{\includegraphics{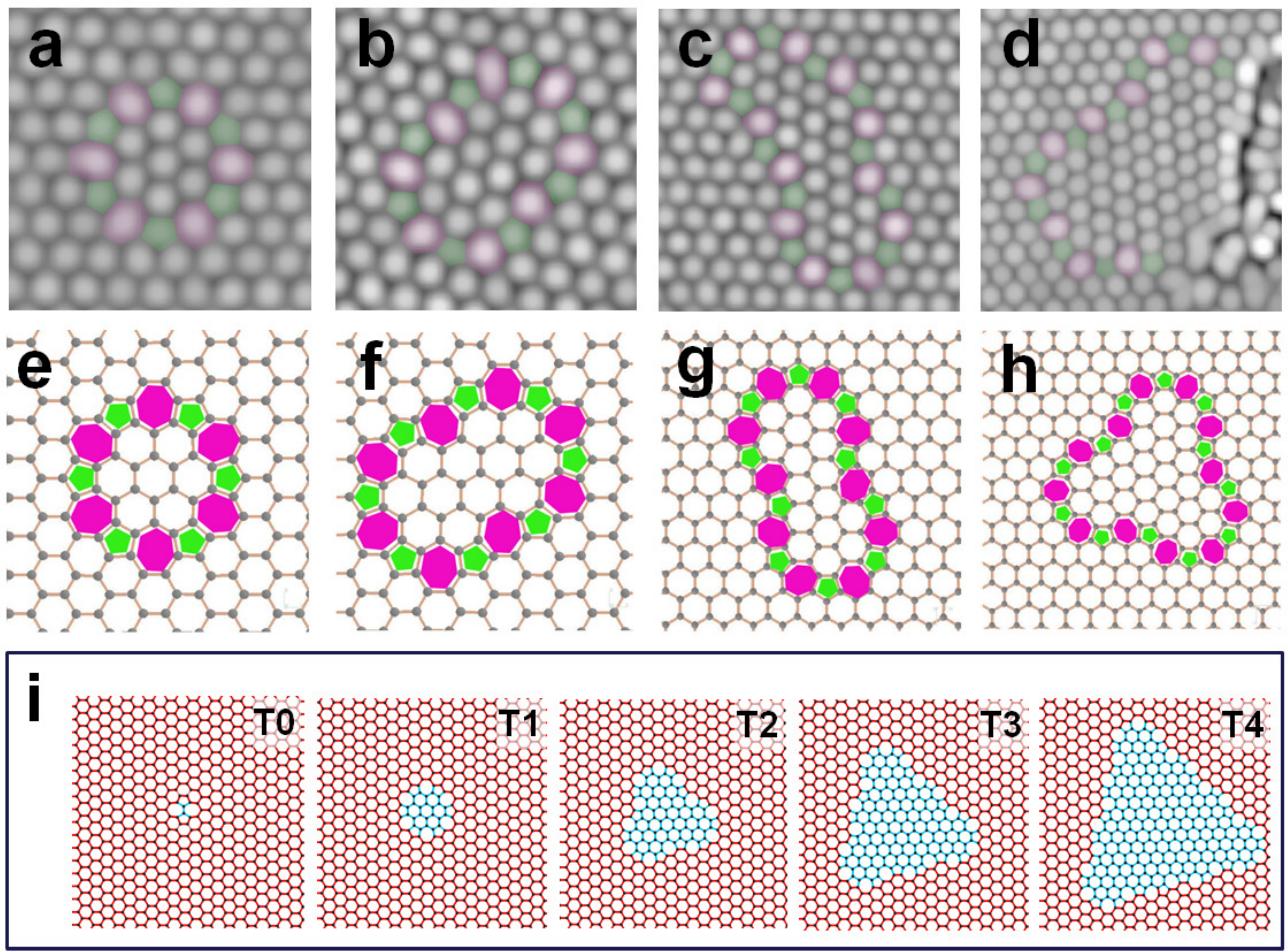}}
  \end{center}
  \caption{
  (a-d)~Experimental \mbox{flower-related} defects of increasing size. 
  (e-h)~DFT calculated \mbox{flower-related} defects of increasing size: (e)~unitary flower T$_1$, (f-g)~double flowers with vacancies and raw, and (h)~triple flower T$_2$. Pentagons and heptagons are highlighted in green and magenta, respectively, in both series.
  (i)~Series of triangular based domain T$_n$ (see text). The different colors correspond to the different orientations of graphene inside (blue) or outside (red) the domain. The formation energy {\it per} pairs of \mbox{5-7} is 2.38, 1.17, 1.21, 1.24 and 1.24~eV for the depicted T$_n$ series.}
  \label{fig:series} 
\end{figure*}

Flower defects were observed few years ago by means of both scanning tunneling microscopy (STM)~\cite{YAN_APL103, Zhao_NL2013} and HRTEM~\cite{LEH_NC4}. 
For graphene over metal, their occurrence was scarce and attributed to a localized dissolution mechanism when the graphene was further heated on its growth metal support~\cite{YAN_APL103}. 
For graphene on SiC, the growth mechanism of flowers remains still unknown~\cite{Zhao_NL2013, CUI_JPCC2017}.
Recently, we have reported the recrystallization of graphene monolayers grown by CVD on platinum~\cite{TYU_CA102}. 
Interestingly, these samples present many \mbox{flower-related} defects as highlighted by the red circles on a large area depicted in \fref{fig:flower}(a) by means of HRTEM. 
These \mbox{flower-related} defects are in fact rotated graphene domains as evidenced by a numerical \mbox{dark-field} image (lower panel in \fref{fig:flower}(a)) realized by selecting one of the two existing orientations (top panel in \fref{fig:flower}(a)). 
Flower defects are observed together with larger domains (closer view are depicted in \fref{fig:series}(a-d)). 
The two main characteristics of these domains are, firstly, a 30$^\circ$ rotation of the inside with respect to the outside graphene matrix, and, secondly, a continuous grain boundary made of alternating \mbox{pentagon-heptagon} \mbox{5-7} pairs. 
The higher stability for such continuous close loops has been predicted by Cockayne \textit{et al.}~\cite{COC_PRB83} and is consistent with our observations. 
These domains were formed during the recrystallization process of \mbox{nano-crystalline} graphene driven by atomic hydrogen, which we have ascribed to enhanced migration of the grain boundary in \cref{TYU_CA102}. 
However, due to  a weaker interaction of graphene with platinum \cite{Wang2016}, the formation of the flower defects on platinum cannot be attributed to a dissolution mechanism as proposed by Yan et al.~\cite{YAN_APL103} for \mbox{rhodium-supported} graphene.

In the present growth conditions, we suggest that the flowers originate from a bulge nucleation mechanism, analogous to that often observed in dynamic crystallization  under strain in geosciences~\cite{RIOS2006, CHAUVE2016}. 
This mechanisms is well documented and it consists of the four main stages \cite{HALFPENNY2006} reported in the bottom panel of \fref{fig:flower}(b). 
One of the main features of this mechanism is the presence of serrated grain boundaries and \textit{left over} grains once the process has completed stage IV \cite{CHAUVE2016}. 
Due to their dynamic character, stages II and III are rarely observed. 
The proposed driving force for the bulging mechanism is the presence of strain during growth and nucleation in the dynamic recrystallization.

By analogy, we propose that the flower defects arise from a similar bulge mechanism in which the flower domain would be the grains \textit{left over} by the dynamic recrystallisation.
In the case of our growth process, strain could be provided by the polycrystallinity of the Pt support. Indeed, such a polycrystalline support is expected to induce a distribution of lattice mismatches with the grown graphene film, which only displays two orientations \cite{TYU_CA102}. 
The first analogy with the bulge mechanism is the presence of a characteristic serrated grain boundary between two graphene grains misoriented by 30$^\circ$. This grain boundary is partially hidden by some amorphous carbon, but can be revealed (yellow line in \fref{fig:flower}(a)) with the numerical \mbox{dark-field} technique illustrated in the lower inset of the figure. 
Such serrated grain boundaries are typical of dynamic recrystallization under strain~\cite{CHAUVE2016} and are observed in a large area of our graphene sample (see also large scale dark field image in figure 3c-d of \cref{TYU_CA102}).
This serrated character is uncommon in CVD-grown graphene, where grains usually have a polygonal geometry.

As a further analogy, we report an almost complete graphene bulge in \fref{fig:flower}(b), corresponding to stage II or III in our sketched mechanism. 
Unfortunately, the right end side grain is hidden by amorphous carbon.
Yet, we can clearly identify the characteristic bottleneck that is about to close thus leaving a \mbox{flower-related} defect in the bottom grain (stage IV). 
It is also worth noticing that the presence of kinks in the grain boundary~\cite{emil2019} is an additional clue for the mobility of the latter. 
In summary, the many \mbox{flower-related} defects present in our recrystallized sample can be seen as the \textit{left over} grains by the dynamics recrystallization with a bulge nucleation.

Although the \mbox{flower-related} defects are supposedly driven by the aforementioned kinetic process, we have evaluated their stability by calculating their formation energy for increasing size by means of DFT. 
This is done by using the flower defect as a stencil for the shape of the rotated domain inside the graphene matrix. 
The bare calculated double flower defect is depicted in \fref{fig:series}(g). 
The interesting double flower with two vacancies that is observed in our sample is found to be more stable than the full double flower (\fref{fig:series}(b) and (f), respectively) by 1.2~eV. 
It is worth noticing that such double flower defects (with or without vacancies) might be related to the \mbox{conjoined-twin} defect observed by STM~\cite{CUI_JPCC2017} in graphene grown on SiC at high temperatures. 
For three-flowers defects, we find that the most favorable domains present a triangular shape (\fref{fig:series}(h)), as it reduces the number of \mbox{5-7} pairs with respect to a linear arrangement. 
Such triangular domains are indeed observed in our samples (\fref{fig:series}(d)). 
We thus decide to keep this triangular shape while increasing the size of the domain to allow a direct comparison as a function of size. 
The calculated formation energies are 7.0, 14.6, 22.3 and 29.8~eV for the four considered triangular domains, T$_n$, \textit{n} being the number of flower units along an edge of the triangle (see \fref{fig:series}(i)). 
The Stone-Wales \cite{stone_theoretical_1986}, being the smallest rotated domain~\cite{COC_PRB83}, can be considered in the series as T$_0$ and its formation energy is 4.8 eV.
The formation energy for $n > 1$ scales almost linearly with \textit{n} and with the number of \mbox{5-7} pairs, and costs roughly 1.2~eV {\it per} pair. 
So, from a thermodynamic point of view, small and compact domains are expected to be more stable as larger domains imply more \mbox{5-7} dislocation cores. 
The unitary flower T$_1$ is the most stable of the series by few meV (\fref{fig:series}(i)) in agreement with its observation at equilibrium growth conditions~\cite{YAN_APL103, Zhao_NL2013, CUI_JPCC2017}.

In the following sections, only the smallest triangular domains will be considered. The use of the \mbox{Stone-Wales}, T$_0$, the unitary flowers, T$_1$, and the triple flowers, T$_2$, will be evaluated as possible building blocks for tuning the electron and phonon transport in all-graphene devices. 
Although it is beyond the scope of the present work, we indeed envision that our proposition of a bulge mechanism during recrystallisation could guide future process development aiming to tailor the flower defects. 
The obvious control parameters to achieve such a goal are: initial grain size, strain level and temperature.

\section{Transport properties of flowered graphene} \label{Sec:Results}

\subsection{Electron transport properties} \label{Sec:Results_ELEC}
To investigate the electron transport properties of flowered graphene, we consider a \mbox{first-neighbor} \mbox{tight-binding} model for the system and perform simulations based on the \mbox{non-equilibrium} Green's function approach, as described more in detail in \sref{Sec:Methods_TBGF}.

As a model system, we consider an armchair ribbon with width \mbox{$W=50$~nm} and a random distribution of flowers with number densities in the range $\rho=10^{11}-10^{13}$~cm$^{-2}$ over a section of length $L=250$~nm. 
Such a ribbon is large enough to allow the observation of the relevant physics of flowered graphene. 
In order to draw more general conclusions, we consider an ensemble of random disordered configurations, whose number is selected depending on the quantity we are interested in.

\begin{figure}[!t]
 \begin{center}
  \resizebox{\cw}{!}{\includegraphics{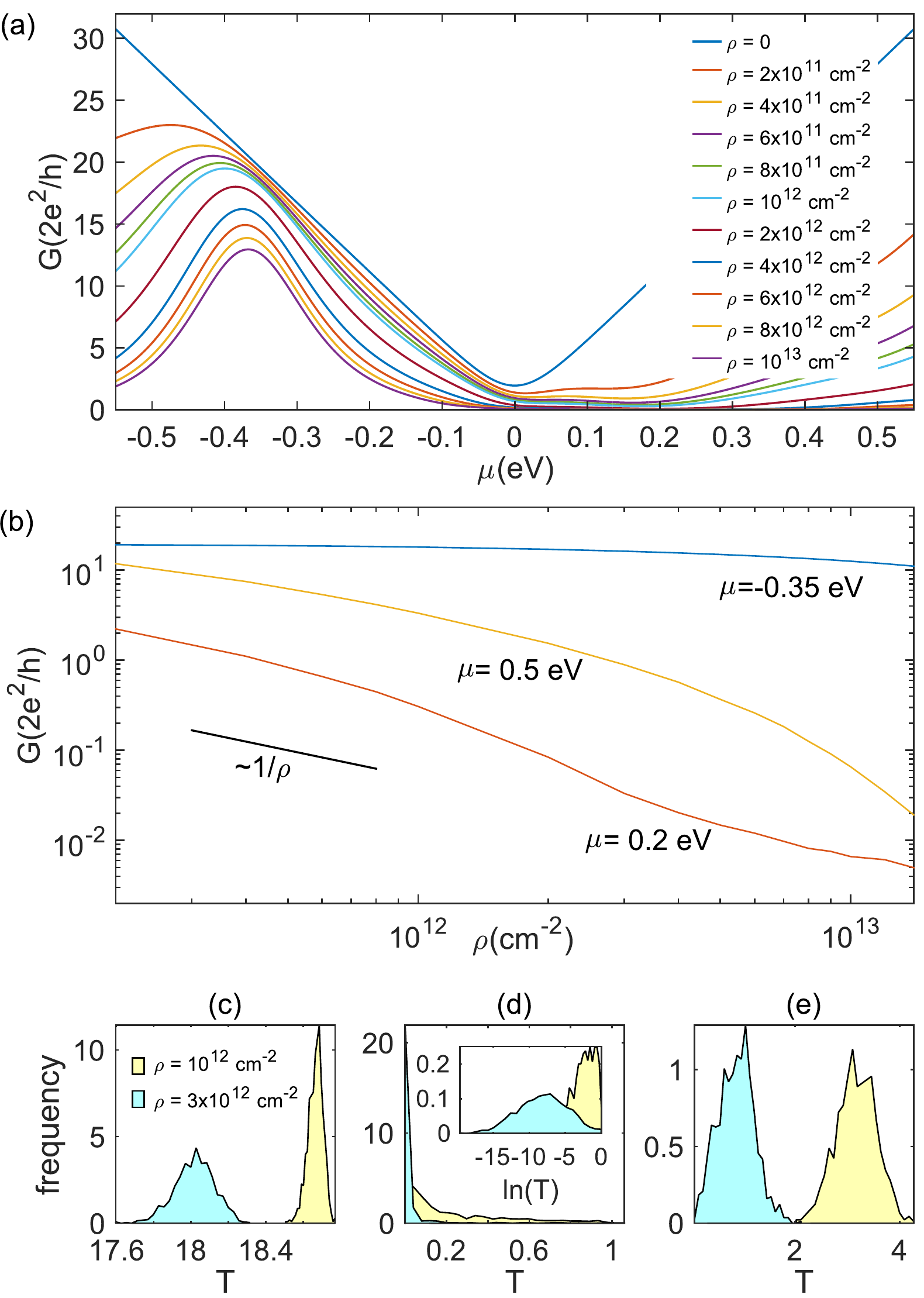}}
 \end{center}
  \caption{(a)~Conductance of monolayer graphene armchair ribbons with width $W=50$~nm and varying concentration $\rho$ of flower defects over a section with length $L=250$~nm at a temperature of 300~K. 
	(b)~Average conductance as a function of the flower concentration for selected electron energies $E=-0.35$~eV, 0.2~eV and 0.5~eV, corresponding to the \mbox{quasi-ballistic}, localized and diffusive regimes. 
	The averaged is performed over 50 random realizations of disorder. 
	(c)~Frequency distribution of the transport coefficients (over 1000 random realizations of disorder) at $E=-0.35$~eV (\mbox{quasi-ballistic} regime) for $\rho=10^{12}$~cm$^{-2}$ and $3\times 10^{12}$~cm$^{-2}$. 
	(d)~Same as (c) at $E=0.2$~eV (localized regime). (e) Same as (c) at $E=0.5$~eV (diffusive regime).}
  \label{fig:electron} 
\end{figure}

\Fref{fig:electron}(a) reports the differential conductance $G$ for pristine and flowered ribbons as a function of the chemical potential $\mu$ and at temperature $T=300$~K. 
The conductance is averaged over an ensemble of 50 random realizations of disorder. We observe a marked \mbox{electron-hole} asymmetry around $\mu=0$, which is the consequence of the sublattice symmetry breaking due to the presence of \mbox{odd-numbered} rings in the flower defects. 
More specifically, the results show three different behaviors depending on the chemical potential. 
For holes close to the charge neutrality point ($-0.4~{\rm eV}<\mu<0$), the conductance is scarcely affected by flowers, and it is close to that of the pristine ribbon, at least for $\rho\lesssim 10^{12}$~cm$^{-2}$. 
For electrons close to the charge neutrality point and when increasing the flower concentration to \mbox{$\rho= 10^{12}$~cm$^{-2}$}, a transport gap develops for chemical potentials up to $\sim 0.2$~eV. 
For higher concentrations, the gap further enlarges.
The opening of an electron transport gap together with the preservation of the hole conductance is a striking result that provides a possible way to turn graphene into a semiconductor.
For energies far away from the charge neutrality point, the conductance decreases more moderately and in a fairly \mbox{electron-hole} symmetric way.

In what follows, we examine these regimes more deeply. 
We select three representative chemical potentials, namely $\mu=-0.35$~eV, $\mu=0.2$~eV and $\mu=0.5$~eV. 
\Fref{fig:electron}(b) shows the corresponding conductance as a function of the flower density. 
For $\mu=-0.35$~eV the conductance does not vary significantly, meaning that the electron scattering induced by flowers is very moderate and that transport is \mbox{quasi-ballistic}. 
This conclusion is supported by the frequency distribution of the transmission coefficient $\mathcal{T}$ at the electron energy $E=-0.35$~eV for $\rho=10^{12}$~cm$^{-2}$ and $3\times 10^{12}$~cm$^{-2}$, see \fref{fig:electron}(c). 
Here, the width of the distribution $\Delta \mathcal{T}$ is much smaller than the average conductance $\langle \mathcal{T} \rangle$, where $\langle ... \rangle$ indicates the average over the ensemble of 1000 disordered configurations. 

For $\mu=0.2$~eV, the conductance as a function of the flower concentration decreases faster than $1/\rho$, even at small densities. 
This suggests that scattering is beyond the dilute limit and impurities couple to give rise to localized states. 
Again, this is confirmed by the frequency distribution for the transmission coefficients, see \fref{fig:electron}(d). 
In this case, the distribution is strongly peaked at low $\mathcal{T}$ with $\Delta \mathcal{T}/\langle \mathcal{T} \rangle>1$. At the same time, the distribution of $\log \mathcal{T}$ is Gaussian, with $\Delta (\log \mathcal{T})/\langle \log \mathcal{T} \rangle<1$. This behavior indicates a localized transport regime~\cite{AVR_MPLB21}. 
A further analysis of the data (not shown here) confirms that the typical scaling law of the localized transport regime holds
\begin{equation}
 \langle \log \mathcal{T}(L)\rangle \ = \ \log\mathcal{T}(L=0) \ - \ L/\xi \ ,
\end{equation}
where the localization length is $\xi\approx 60$~nm for $\rho=10^{12}$~cm$^{-2}$, and $\xi\approx 25$~nm for $\rho=3\times 10^{12}$~cm$^{-2}$. 

Finally, for $\mu=0.5$ eV, the average conductance decreases as $1/\rho$ up to almost $\rho\approx 3\times10^{12}$ cm$^{-2}$, thus suggesting a diffusive transport regime. 
This is confirmed by the Gaussian distribution of the transport coefficients (with $\Delta \mathcal{T}/\langle \mathcal{T} \rangle < 1$) as reported in \fref{fig:electron}(e), and by the scaling law
\begin{equation}
 \langle \mathcal{T}(L)\rangle \ = \ \mathcal{T}(L=0) \ \Frac{\ell}{L+\ell} \ ,
\end{equation}
where the mean free path is $\ell\approx 35$~nm for $\rho=10^{12}$~cm$^{-2}$. 
At higher concentrations, the average conductance decays faster than $1/\rho$, thus indicating a transition to the localized regime and explaining the widening of the transport gap at higher energies, as observed in \fref{fig:electron}(a). 
Indeed, for $\rho=3\times 10^{12}$ cm$^{-2}$ the transport regime is in between localized and diffusive, with $\xi\approx 147$~nm and $\ell\approx 4$~nm.

\begin{figure*}[!t]
\begin{center}
\resizebox{\textwidth}{!}{\includegraphics{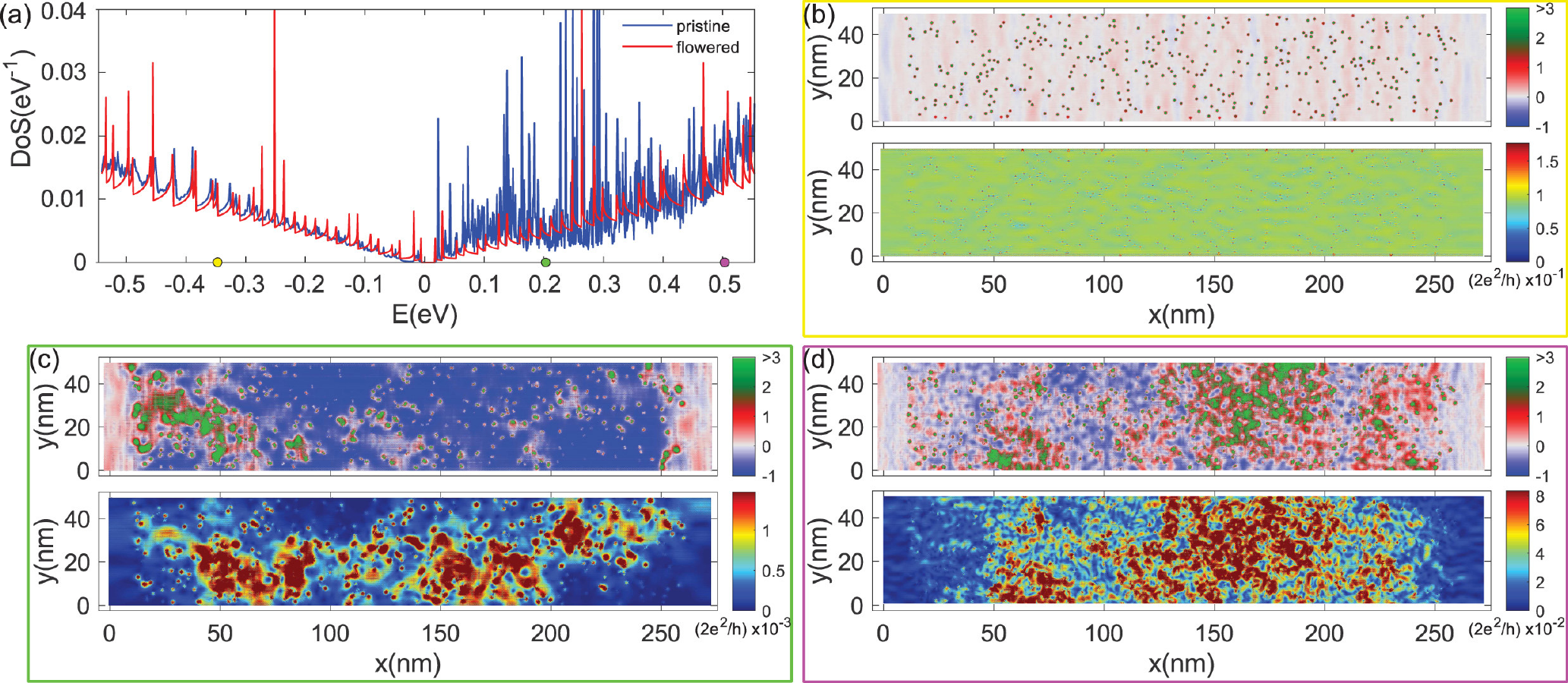}}
\end{center}
\caption{
(a)~Density of states {\it per} atom as a function of the energy for pristine and flowered ($\rho=3\times 10^{12}$ cm$^{-2}$) \mbox{50-nm-wide} ribbons. 
(b)~Local variation of the density of states (top) and local spectral current intensity (bottom) at $E=-0.35$~eV, indicated by a yellow dot in (a). (c)~The same as (b) at $E=0.2$~eV, indicated by a green dot in (a). 
(d)~Same as (b) at $E=0.5$~eV, indicated by a magenta dot in (a).}
\label{fig:maps} 
\end{figure*}

Further physical insight into these results can be gained by looking at the density of states (DoS). 
\Fref{fig:maps}(a) shows the DoS as a function of the energy for a \mbox{50-nm-wide} ribbon in the absence and in the presence of flowers with a density $\rho=3\times 10^{12}$~cm$^{-2}$. 
A single realization of disorder is considered.
For holes, the density of states is only marginally affected by flowers. This explains why the hole conductance is very robust against the presence of flowers. 
On the contrary, for electrons, the density of states is strongly modified with respect to the pristine case. 
The appearance of many additional spikes indicates the formation of (more or less localized) states, which explains why the electron conductance is more strongly affected by flowers. 
To further substantiate this picture, we analyze the local density of states and the local distribution of the spectral currents at the three representative energies $E=-0.35$~eV, $E=0.2$~eV and $E=0.5$~eV.
The top panels of \ffref{fig:maps}(b-d) illustrate the variation of the local density of states of the disordered system with respect to the pristine system. 
The blue regions correspond to a depletion of states (the deep blue indicates a variation -100\%, i.e. a local DoS close to 0). 
The red regions indicate an increase of the local DoS up to 100\%. 
Finally, the DoS is very high in the green regions. 
The bottom panels of \ffref{fig:maps}(b-d) illustrate the local spectral electron current that flows along the ribbon, varying from blue (low currents) to red (high currents). 
In the quasiballistic regime, see \fref{fig:maps}(b), the DoS varies more significantly on the flowers, but very weakly in the regions between them. 
As a consequence, transport is scarcely affected, as we can see from the fairly homogeneous distribution of the spectral currents.
In the localized regime, see \fref{fig:maps}(c), the DoS is strongly enhanced around the flowers and suppressed in the rest of the system, thus reducing available states for transport. 
Therefore, the electron propagation is made difficult and indeed strongly suppressed. 
This is clearly seen from the local distribution of spectral currents, where electrons appear to circulate in the \mbox{high-DoS} regions without being able to cross the \mbox{low-DoS} areas.
Finally, in the diffusive regime, see \fref{fig:maps}(d), irregular variations of the DoS in the region between the defects are present, thus introducing scattering for electrons. 
Accordingly, the spectral current is fragmented. 
Note that a periodic arrangement of the defects to form a single and ordered grain boundary would, on the contrary, give rise to extended states with enhanced conductance~\cite{Ma2014}. 
For the sake of completeness, we add that in the case of larger flowers (T$_2$, T$_3$ and T$_4$) the behavior of the three transport regimes is very similar. 
From the results (not show here), we find that the energy ranges where the transport regime is \mbox{quasi-ballistic} or localized slightly narrow with the flower size, while the diffusive regimes show a transmission coefficient that scales as the flower perimeter, i.e. proportionally to the number of pentagon/heptagon carbon rings.

\begin{figure}[t]
  \begin{center}
   \resizebox{\cw}{!}{\includegraphics{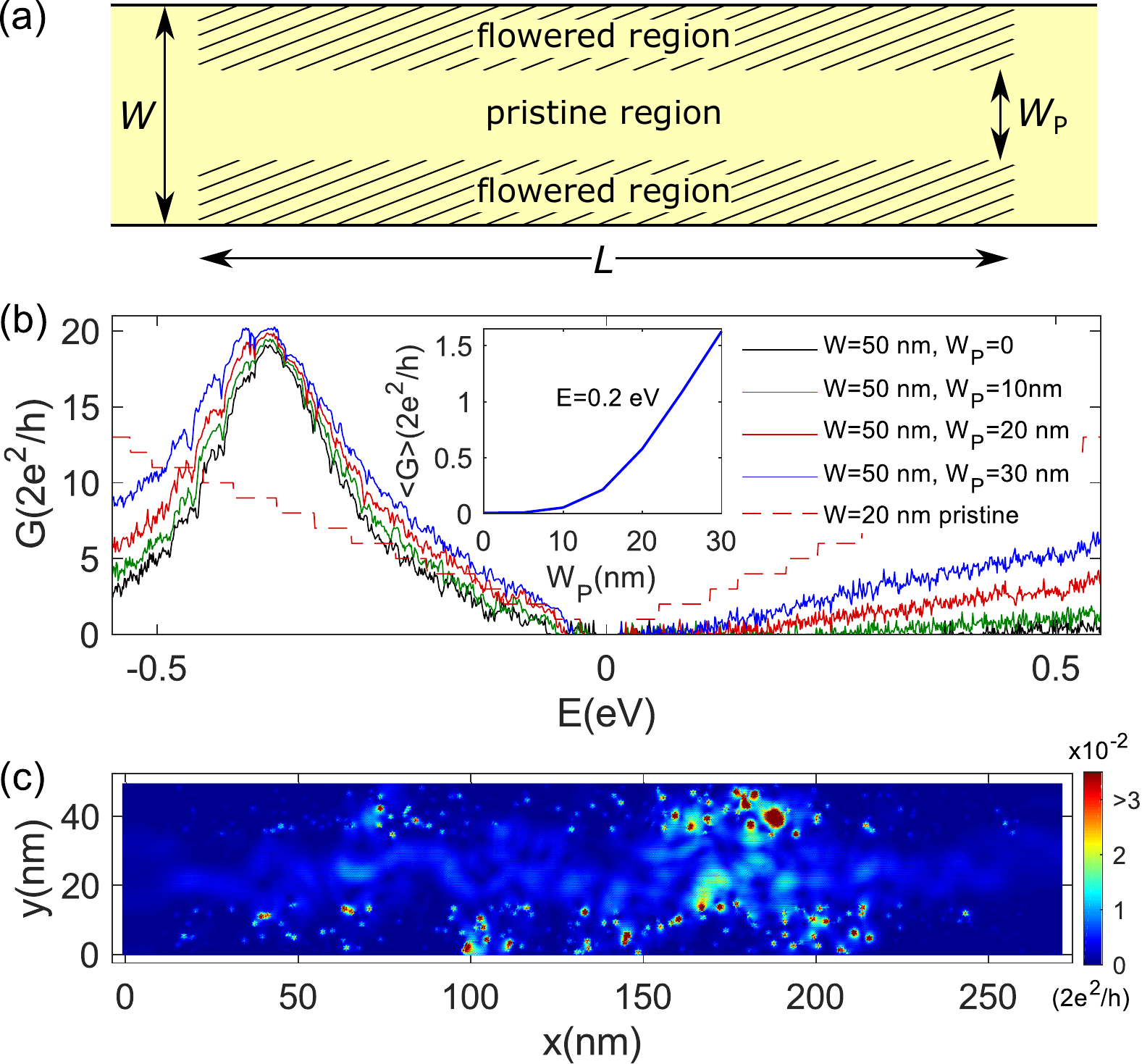}}
  \end{center}
\caption{(a)~Sketch of the simulated system constituted of a ribbon with width $W$ with a flowered region of length $L$ and a pristine flower-free channel with width $W_{\rm P}$. 
(b)~\mbox{Zero-temperature} conductance for a ribbon of width $W=50$~nm, with flower impurities distributed over a length $L=250$~nm with concentration $\rho=5\times 10^{12}$~cm$^{-2}$ and with a flower-free central stripe of width $W_{\rm P}$ in the range [0,30]~nm. 
Inset:~Average \mbox{zero-temperature} conductance at $E=0.2$~eV and as a function of the flower-free channel width $W_{\rm P}$. 
An ensemble of 1000 random configurations was considered.
(c)~Spectral current distribution for $W_{\rm P}=5$~nm at $E=0.2$~eV.}
  \label{fig:stripes} 
\end{figure}

As shown above, for high flower density, the asymmetric transport gap acts as a filter that backscatters electrons and lets holes flow. 
This phenomenon could be exploited to open conductive paths for electrons in graphene by selectively placing flowers in certain regions. 
Electrons ($E>0$) will be free to flow in the clean regions, while they will hardly penetrate the flowered regions. 
We test this idea by simulating electron transport in our \mbox{$W=50$~nm} wide armchair ribbon with a flower concentration $\rho=5\times 10^{12}$~cm$^{-2}$ (over a length $L=250$~nm) and a flower-free pristine region along the ribbon axis with different widths $W_{\rm P}$ varying in the range [0,30]~nm, see \fref{fig:stripes}(a). 
Note that the edges of the flower-free regions are rough, since they are defined by the irregular interface with the flowered region. 
The resulting conductance is reported in the main panel of \fref{fig:stripes}b for individual disordered configurations.
We observe that it decreases roughly linearly with $W_{\rm P}$, as expected, but it is far from being quantized. 
Indeed, the conductance of a $W_{\rm P}=20$~nm ribbon defined by flower defects is about one third that of the equivalent pristine ribbon of width $W=W_{\rm P}=20$~nm, see the dashed line in \fref{fig:stripes}(b).
This is due to the fact that the current can penetrate the disordered flowered regions that define the edges of the clean channel, which results in something analogous to the effect of edge roughness~\cite{MUC_PRB79,DUB_ACSN4,SAL_PRB83,IHN_PRB88}.

To better elucidate this behavior, \fref{fig:stripes}(c) shows the spatial distribution of spectral currents at $E=0.2$~eV and for $W_{\rm P}=20$~nm. 
We can clearly observe that the electron flow is almost confined in the flower-free channel. 
However, electrons are scattered by the rough edges of the channel and can partly penetrate the flowered region. 

The impact of roughness is expected to be more important for narrow ribbons, where the disordered edge area is large and comparable to the clean inner area. 
This is demonstrated by the inset of \fref{fig:stripes}(b), which reports the conductance of the system at $E=0.2$~eV as a function of the pristine channel width $W_{\rm P}$ and averaged over 1000 disordered configurations. 
We observe that the transmission is strongly suppressed up to about $W_{\rm P}=15$~nm, and then it starts increasing more significantly when increasing $W_{\rm P}$.

We can conclude that the engineering of flower defects could enable their exploitation to create conductive paths in 2D graphene for future \mbox{all-graphene} circuits. 
However, due to the intrinsically rough definition of the edges, the conductance of the conductive paths, especially if narrow, would be lower than the corresponding pristine system, with large \mbox{defect-dependent} variability and possible residual current penetration in the disordered region. 

\subsection{Thermal transport properties} \label{Sec:Results_THER}
Next, we study the effect on flower defects and related crystallographic imperfections on the thermal conductivity of graphene from a solution of the \mbox{Boltzmann-Peierls} transport equation for phonons. 
Under the \mbox{relaxation-time} approximation, the thermal conductivity tensor can be expressed as~\cite{cpc_2014}:

\begin{equation}
  \kappa^{\left(\mu\nu\right)} = \frac{1}{k_BT^2V}\sum\limits_\lambda n_{\lambda} \left(n_{\lambda} + 1\right)\left(\hbar \omega_{\lambda}\right)^2 v_{\lambda}^{\left(\mu\right)}v_{\lambda}^{\left(\nu\right)} \tau_{\lambda},
  \label{eqn:kappa}
\end{equation}

\noindent where $\mu$ and $\nu$ are Cartesian indices, $\lambda$ is a combined mode index denoting a phonon wave vector $\boldsymbol{q}$ and a phonon branch index $\alpha$, $V$ is the volume of the unit cell, and $\omega_{\lambda}$, $\boldsymbol{v}_{\lambda}$ and $\tau_{\lambda}$ are the angular frequency, group velocity and relaxation time of mode $\lambda$, respectively. 
The sum $\sum_{\lambda}$ stands for the combination of a literal sum over branches and an average over the Brillouin zone. 
Due to the symmetries of pristine graphene, the tensor $\boldsymbol{\kappa}$ is isotropic and can be treated as a scalar $\kappa$. 
Addition of defects does not change this fact as long as their orientations are symmetrically distributed.

We employ \mbox{first-principles} calculations to characterize the phonon spectrum and the intrinsic scattering rates of graphene~\cite{cpc_2014}. 
We then use Green's function methods as implemented in the almaBTE package~\cite{katre_unraveling_2016,Wang_PRB17,almaBTE} to obtain the elastic contribution of the defects to the total $\tau_{\lambda}^{-1}$ for each mode. We treat the perturbation introduced into the interatomic force constants by the defects by using a semiempirical potential specifically designed for thermal conductivity calculations~\cite{lindsay_optimized_2010}. This combination of \mbox{first-principles} calculations for inelastic scattering and semiempirical potentials for elastic scattering has been shown to afford \mbox{{\it ab-initio}-like} accuracy in previous studies~\cite{dislocations2019}. All relevant details are provided in \sref{Sec:Methods_PH}.

We compare the effects of three different kinds of crystallographic imperfections: \mbox{Stone-Wales} (T$_0$)~\cite{stone_theoretical_1986}, unitary flower (T$_1$) and triple flower (T$_2$) defects. Conceptually speaking, all three can be considered as particular cases of topological defects comprising a perimeter of pentagons and heptagons that enclose a finite core of hexagonal graphene cells with lattice axes at $90^\circ$ with respect to those of the main lattice. 
In the T$_0$ case, there is no core and the perimeter contains two heptagons and two pentagons. 
On the other hand, each T$_1$ and T$_2$ defect contains $19$ and $52$ graphene rings, respectively, considering both their cores and their perimeters.

\begin{figure}[!t]
  \begin{center}
    \resizebox{\cw}{!}{\includegraphics{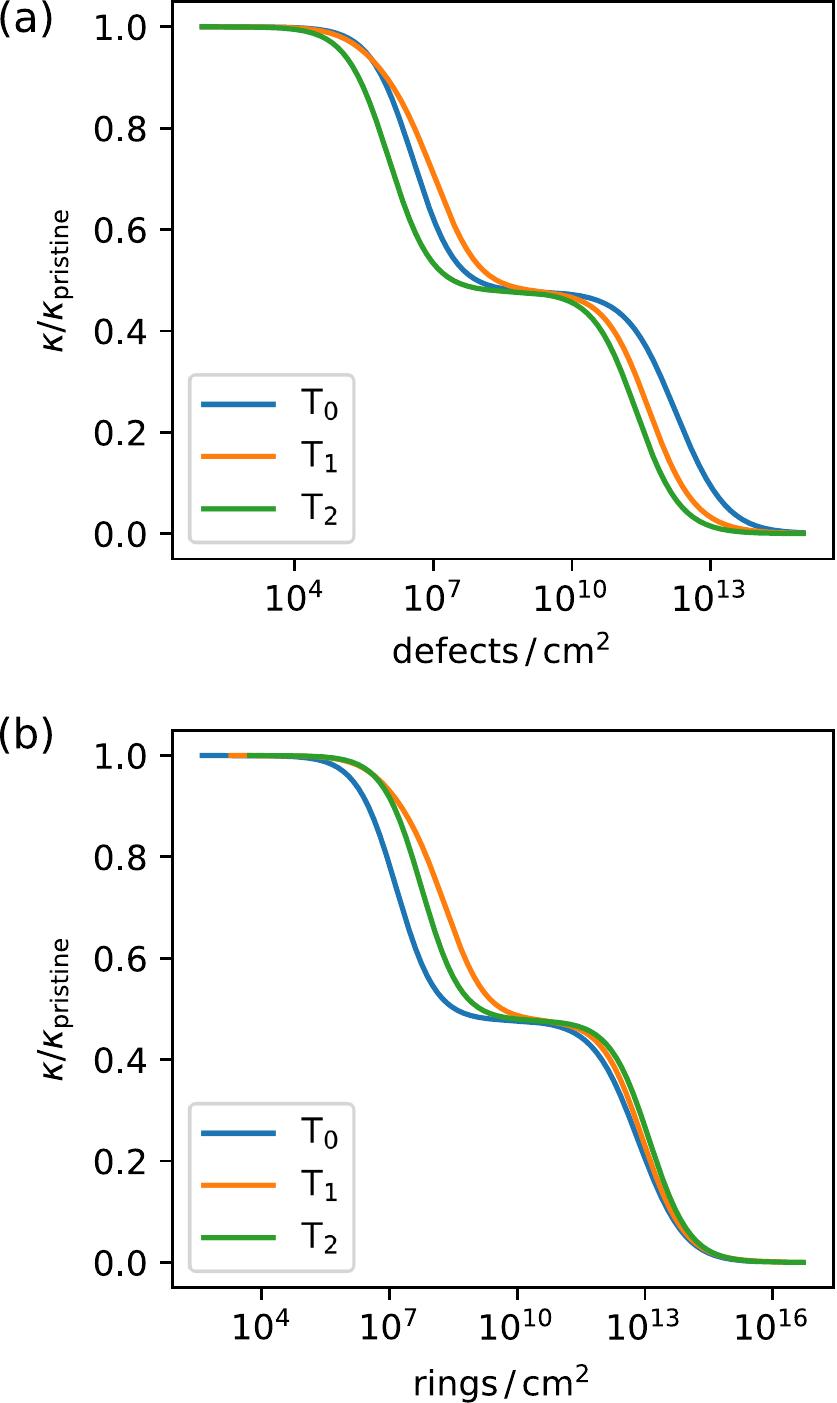}}
  \end{center}
  \caption{\mbox{Room-temperature} thermal conductivity of a monolayer graphene sample with varying concentration of T$_0$, T$_1$ or T$_2$ defects as a function of (a)~the number density of defects or (b)~the number density of rings contained in the defects.}
  \label{fig:thermal_conductivity}
\end{figure}

\Fref{fig:thermal_conductivity} shows how the \mbox{room-temperature} thermal conductivity of graphene is reduced by different concentrations of such defects. The top and bottom panels represent the same data, but with concentration quantified in two different ways. The horizontal axis in the top panel represents the numerical density of defects regardless of their size, while in the bottom panel it represents the numerical density of graphene rings making up the defects.

\begin{figure}[!t]
    \begin{center}
    \resizebox{\cw}{!}{\includegraphics{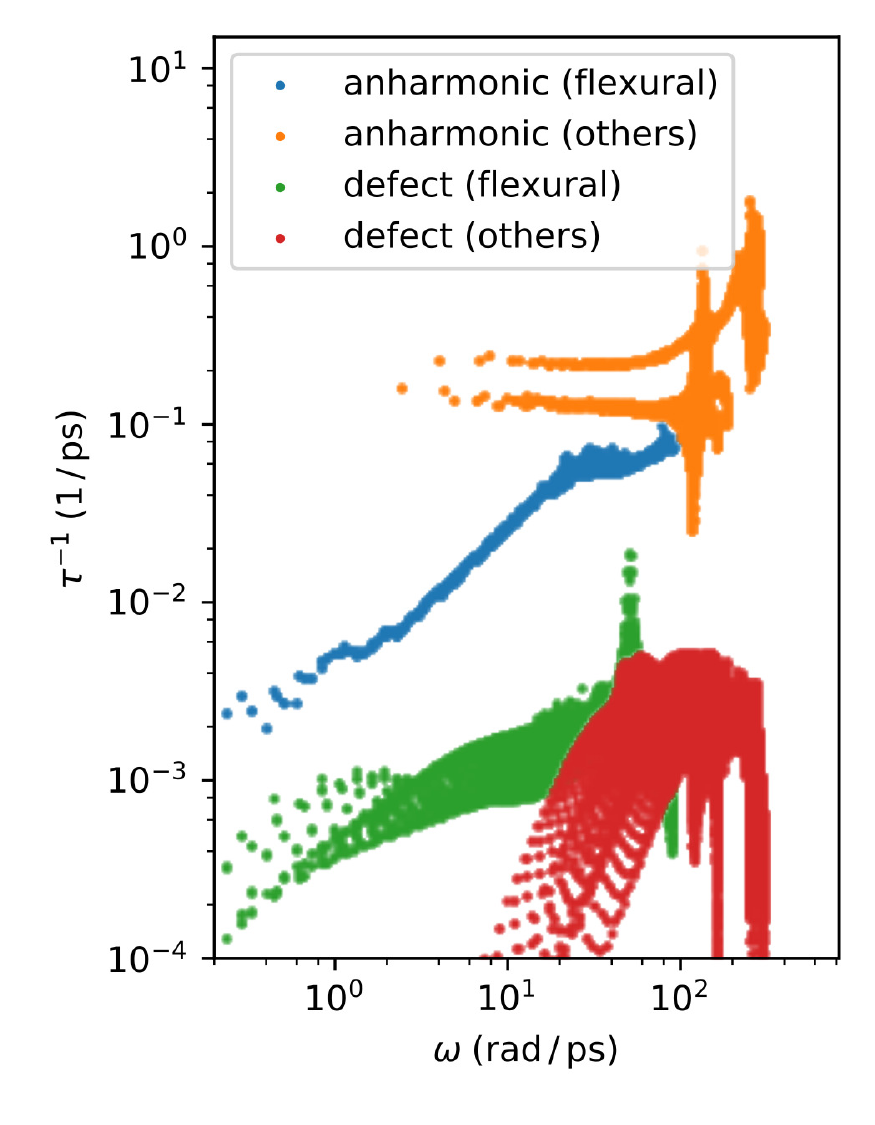}}
    \end{center}
    \caption{Computed intrinsic anharmonic scattering rates for phonons in graphene at \mbox{$T=300$~K} alongside the elastic scattering rates corresponding to a number density of T$_1$ defects $\rho_0 = 10^{10}\;\mathrm{cm^{-2}}$.}
    \label{fig:phonon_scattering}
\end{figure}

A salient feature of all curves in \fref{fig:thermal_conductivity} is the plateau at intermediate concentrations, which interrupts the otherwise monotonic decrease of $\kappa$ with increasing defect concentration. 
This turns out to be a manifestation of the peculiar physics of phonons in graphene and the singular importance of the flexural branch for transport. As illustrated in \fref{fig:phonon_scattering}, the intrinsic scattering rates for the flexural branch are substantially smaller than those of the other branches, and show a steeper dependence on the frequency for low energies. This is a result of the quadratic dispersion of that branch~\cite{mrl2016} and the aforementioned symmetry-induced selection rule for \mbox{three-phonon} scattering processes~\cite{lindsay_flexural_2010}. 
Moreover, the elastic scattering rates for the flexural branch are higher than those for the others and decay much more slowly when the frequency approaches zero, a phenomenon also observed in other 2D calculations~\cite{Wang_PRB17}. Hence, as more defects are introduced, the elastic contribution to scattering rates first becomes comparable to the intrinsic scattering rates for this flexural branch, thus significantly reducing its contribution to thermal transport. 
The presence of an interval of concentrations where elastic scattering is strong enough to drastically suppress the thermal transport by the \mbox{low-frequency} region of the flexural branch, but not enough to be relevant for all other branches, is the reason for the plateaus of \fref{fig:thermal_conductivity}.

Another conclusion to be drawn from the thermal conductivity vs. concentration curves concerns the scattering efficiency of each kind of defect. 
Elastic phonon scattering comes about as a result of a breakdown of strict periodicity in the crystal, and hence it could be expected that larger crystallographic imperfections lead to stronger phonon scattering. 
However, it can be seen from the top panel of \fref{fig:thermal_conductivity} that the situation is more nuanced. 
In the \mbox{high-concentration} regime, where elastic scattering significantly affects all phonon branches, the order is exactly as would be expected from this simplistic argument, but for concentrations below the conductivity plateau each T$_0$ defect scatters phonons more intensely than a T$_1$ defect in spite of its smaller size. 
This goes to show that the scattering intensity depends on the fine details of the coupling between the propagating vibrational modes in the crystal and the more localized ones around the defect, which can still be significantly more extended than the defect itself. The point becomes even clearer in \fref{fig:thermal_conductivity}(b), where the horizontal axis represents the fraction of the area of the graphene layer covered by defects and which reveals that, on a \mbox{hexagon-by-hexagon} basis, the \mbox{Stone-Wales} defect T$_0$ is more efficient at scattering phonons than any of the other two. 
The effect is comparable to classical results on phonon scattering by spherical nanoparticles~\cite{kim_phonon_2006}, which show a transition from the Rayleigh to the geometrical limits (with the corresponding reduction in scattering efficiency {\it per} unit volume of the defect) as the particle size is increased. In other words, the agglomeration of \mbox{defect-laden} unit cells reduces their effect on phonon propagation as compared with the same number of independent scatterers. Interestingly, the triple flower (T$_2$) is still a more efficient phonon scatterer than T$_1$, maybe because of the breakdown of \mbox{in-plane} inversion symmetry it entails.

We also note that for the kind of concentrations considered in previous sections ($\rho=10^{11}-10^{13}$ cm$^{-2}$), defects reduce the thermal conductivity of graphene by up to two orders of magnitude by dramatically depressing the contributions from all the acoustic branches. 
Such an effect can be interesting with a view to thermoelectric energy generation.

\section{Conclusion} \label{Sec:Conclusions}

To summarize, we investigated the structural and transport properties of flowered graphene. Our study of the structural properties of flower defects by \mbox{high-resolution} transmission electron microscopy and {\it ab initio} simulation allowed us to propose a new mechanism underlying their growth. 
Our findings represent a first step toward the future development of experimental techniques to control the flower density and position by controlling the bulge nucleation mechanism that we identified during the recrystallization process. 
The simulation of electron transport in large flowered graphene samples revealed a strongly asymmetric conductance, with a large transmission coefficient for holes and the development of a transport gap for electrons. 
Such an asymmetry originates from the presence of \mbox{odd-numbered} carbon rings, which break the sublattice symmetry of graphene. 
We analyzed the resulting \mbox{quasi-ballistic}, diffusive and localized transport regimes and described them in terms of local density of states, local spectral current distribution and frequency distribution of the transmission coefficient for a large ensemble of disordered configurations. 
Finally, we reported a strong reduction in \mbox{room-temperature} thermal conductivity due to the insertion of flower defects, as well as the related \mbox{Stone-Wales} defects and triple flowers, in the graphene sheet. 
The dependence of the calculated conductivity on defect density shows that the concentration required to drastically impede heat transport by flexural phonons is rather modest, around $10^{10}\;\mathrm{cm^{-2}}$. 
Higher concentrations ($~10^{13}\;\mathrm{cm^{-2}}$) can affect all phonon branches and reduce thermal conductivity by up to two orders of magnitude. 
However, flower defects are less effective phonon scatters than \mbox{Stone-Wales} defects covering the same area.

In conclusion, on one hand, our results are useful for characterizing the transport properties of defective CVD graphene, which is among the most promising \mbox{large-scale} growth techniques for graphene. 
On the other hand, our findings may also be exploited for more practical applications, as energy filtering or creation of conductive paths in graphene. 

\section{Experimental and numerical methods} \label{Sec:Methods}

\subsection{High-resolution transmission electron microscopy} \label{Sec:Methods_TEM}
CVD graphene was synthesized on Pt thin layer substrate \cite{TYU_CA102} and transferred onto the HRTEM grid. 
HRTEM experiments were performed using a double aberration-corrected FEI Titan Ultimate operated at 80~kV. 
The monochromator was excited to reduce the energy spread of the electron beam to 0.15~eV. 
HRTEM images were acquired using a CCD Ultra Scan camera and treated by \mbox{low-pass} filtering based on FFT and numerical \mbox{max-filtering} in order to identify pentagons and heptagons formed in the graphene hexagonal structure.  

\subsection{{\it Ab initio} density functional approach} \label{Sec:Methods_DFT}
Calculations were performed with the BigDFT software~\cite{genovese2008} in a supercell that guarantees negligible elastic interactions between the periodic defects~\cite{machado2012, POC_PRB2014}. 
Due to the large flower clusters we considered, an orthorhombic supercell with 836 atoms was needed. 
Geometries were considered optimized when the forces on atoms were less than 15 meV/\AA. 
We used a \mbox{Perdew-Burke-Ernzerhof} (PBE) \mbox{exchange-correlation} functional~\cite{vasp_pbe_1} together with Hartwigsen-Goedecker-Hutter pseudopotentials~\cite{newHGH2013} and only the $\Gamma$ point. 
The formation energy was calculated directly with respect to bare graphene, as flower defects do not contain vacancies or interstitials. 

\subsection{Tight-binding Hamiltonian and Green's function approach for electron transport simulations} \label{Sec:Methods_TBGF}
To describe graphene and flowers we adopted a tight-binding Hamiltonian with a single $p_z$ orbital {\it per} carbon atom and hopping parameter $t=-2.7$~eV. The Hamiltonian reads
\begin{equation}
  H \ = \ \sum_{<ij>} t \ c_i^\dag \ c_j \ ,
\end{equation}
where $c_i^\dag$ and $c_i$ are the creation and annihilation operators for electrons on the carbon identified by the index $i$, and $<...>$ indicates couples of first neighbor atoms. 
From the DFT calculations, it turned out that the strain of the \mbox{C-C} distance ($<2\%$) was very localized in the region of the flowers. 
In a first approximation, we thus disregarded the corresponding variation of the hopping parameter.

The electron transport properties were simulated with the Green's function approach in the \mbox{two-terminal} configuration, i.e. with the flowered ribbon connected to electron reservoirs by two \mbox{semi-infinite} pristine graphene ribbons of the same width. 
The transmission coefficient can be expressed in a Landauer form as
\begin{equation}
  \mathcal{T}(E) \ = \ {\rm Tr} \left[ \ G^R(E) \ \Gamma^{\rm S}(E) \ G^A(E) \  \Gamma^{\rm D }(E)\  \right] \ ,
\end{equation}
where $E$ is the energy of the injected electrons, $G^{R/A}$ are the retarded and advanced Green's functions, and $\Gamma^{\rm S/D}$ are the linewidth functions of source and drain contacts. 
The finite temperature differential conductance $G$ was obtained from the transmission coefficients as
\begin{equation}
	G(\mu,T) \ = \ \Frac{2e^2}{h} \ \int \ \Frac{\mathcal{T}(E)}{4 k_{\rm B} T\cosh^{2}\left(\Frac{E-\mu}{2 k_{\rm B} T}\right)} \ dE \ ,
\end{equation}
where $T$ is the temperature, $\mu$ is the chemical potential, $k_{\rm B}$ is the Boltzmann constant, and $2e^2/h$ is the \mbox{spin-degenerate} conductance quantum.

Within the same formalism, we obtained the local density of states (proportional to the imaginary part of the retarded Green's function) and the spatial distribution of spectral currents~\cite{CRE_PRB68}. 

\subsection{Phonon transport simulations} \label{Sec:Methods_PH}
Starting from the relaxed coordinates of pristine graphene, we employed Phonopy~\cite{phonopy} to generate a minimal set of displaced $9\times 9$ supercell configurations. 
We obtained the forces on all atoms in those configurations the DFT package VASP~\cite{vasp_general_4} with \mbox{projector-augmented-wave} datasets~\cite{vasp_paw_1, vasp_paw_2}, the PBE approximation to exchange and correlation~\cite{vasp_pbe_1}, a \mbox{plane-wave} cutoff of $520\;\mathrm{eV}$ and a simulation box with a height of $17\;\mbox{\normalfont\AA}$ to avoid spurious interactions between periodic images of the graphene layers. 
From these forces, we rebuilt the harmonic force constants needed to obtain the phonon spectrum of graphene. 
We followed a similar procedure using a minimal set of displaced configurations generated using thirdorder.py~\cite{cpc_2014} to obtain the \mbox{third-order} force constants for the supercell. 
As described in \cref{cpc_2014}, these are the ingredients required to obtain the intrinsic scattering rates in the material. 
The 2D causal phonon Green's function of pristine graphene was obtained as described in \cref{Wang_PRB17} using a dense $163\times 163$ grid.

We then obtained minimized configurations of the \mbox{Stone-Wales}, flower and triple flower defects embedded in $9\times 9$ graphene supercells using a variation on the Tersoff potential specifically optimized for \mbox{thermal-conductivity} calculations in graphene~\cite{lindsay_optimized_2010}. 
Using an explicit parameterization of the potential energy allowed us to extract all of its first and second derivatives with respect to positions more efficiently by means of automatic differentiation. 
From the difference in the second derivatives with respect to pristine graphene we extracted a perturbation matrix, which we combined with the phonon Green's function to compute the elastic scattering rates due to the defect following the procedure outlined in \ccref{katre_unraveling_2016, almaBTE, Katre_PRL17}.

Calculations of phonon frequencies, group velocities, intrinsic and extrinsic scattering rates, and Green's functions were carried out using the almaBTE software package~\cite{almaBTE}. 
All thermal conductivity calculations used a $150\times 150$ regular grid in reciprocal space.

\section*{Acknowledgments}
The authors acknowledge the support from the European Union's Horizon 2020 Research and Innovation Programme [grant no. 645776 (ALMA)]. 
Computing time has been provided by the national \mbox{GENCI-CINES} supercomputing centers under contract 6194.


\begin{thebibliography}{94}%
\makeatletter
\providecommand \@ifxundefined [1]{%
 \@ifx{#1\undefined}
}%
\providecommand \@ifnum [1]{%
 \ifnum #1\expandafter \@firstoftwo
 \else \expandafter \@secondoftwo
 \fi
}%
\providecommand \@ifx [1]{%
 \ifx #1\expandafter \@firstoftwo
 \else \expandafter \@secondoftwo
 \fi
}%
\providecommand \natexlab [1]{#1}%
\providecommand \enquote  [1]{``#1''}%
\providecommand \bibnamefont  [1]{#1}%
\providecommand \bibfnamefont [1]{#1}%
\providecommand \citenamefont [1]{#1}%
\providecommand \href@noop [0]{\@secondoftwo}%
\providecommand \href [0]{\begingroup \@sanitize@url \@href}%
\providecommand \@href[1]{\@@startlink{#1}\@@href}%
\providecommand \@@href[1]{\endgroup#1\@@endlink}%
\providecommand \@sanitize@url [0]{\catcode `\\12\catcode `\$12\catcode
  `\&12\catcode `\#12\catcode `\^12\catcode `\_12\catcode `\%12\relax}%
\providecommand \@@startlink[1]{}%
\providecommand \@@endlink[0]{}%
\providecommand \url  [0]{\begingroup\@sanitize@url \@url }%
\providecommand \@url [1]{\endgroup\@href {#1}{\urlprefix }}%
\providecommand \urlprefix  [0]{URL }%
\providecommand \Eprint [0]{\href }%
\providecommand \doibase [0]{http://dx.doi.org/}%
\providecommand \selectlanguage [0]{\@gobble}%
\providecommand \bibinfo  [0]{\@secondoftwo}%
\providecommand \bibfield  [0]{\@secondoftwo}%
\providecommand \translation [1]{[#1]}%
\providecommand \BibitemOpen [0]{}%
\providecommand \bibitemStop [0]{}%
\providecommand \bibitemNoStop [0]{.\EOS\space}%
\providecommand \EOS [0]{\spacefactor3000\relax}%
\providecommand \BibitemShut  [1]{\csname bibitem#1\endcsname}%
\let\auto@bib@innerbib\@empty
%</preamble>
\bibitem [{\citenamefont {Novoselov}\ \emph {et~al.}(2004)\citenamefont
  {Novoselov}, \citenamefont {Geim}, \citenamefont {Morozov}, \citenamefont
  {Jiang}, \citenamefont {Zhang}, \citenamefont {Dubonos}, \citenamefont
  {Grigorieva},\ and\ \citenamefont {Firsov}}]{NOV_SCI306}%
  \BibitemOpen
  \bibfield  {author} {\bibinfo {author} {\bibfnamefont {K.~S.}\ \bibnamefont
  {Novoselov}}, \bibinfo {author} {\bibfnamefont {A.~K.}\ \bibnamefont {Geim}},
  \bibinfo {author} {\bibfnamefont {S.~V.}\ \bibnamefont {Morozov}}, \bibinfo
  {author} {\bibfnamefont {D.}~\bibnamefont {Jiang}}, \bibinfo {author}
  {\bibfnamefont {Y.}~\bibnamefont {Zhang}}, \bibinfo {author} {\bibfnamefont
  {S.~V.}\ \bibnamefont {Dubonos}}, \bibinfo {author} {\bibfnamefont {I.~V.}\
  \bibnamefont {Grigorieva}}, \ and\ \bibinfo {author} {\bibfnamefont {A.~A.}\
  \bibnamefont {Firsov}},\ }\href {\doibase 10.1126/science.1102896} {\bibfield
   {journal} {\bibinfo  {journal} {Science}\ }\textbf {\bibinfo {volume}
  {306}},\ \bibinfo {pages} {666} (\bibinfo {year} {2004})}\BibitemShut
  {NoStop}%
\bibitem [{\citenamefont {Ferrari}\ \emph {et~al.}(2015)\citenamefont
  {Ferrari}, \citenamefont {Bonaccorso}, \citenamefont {Fal'ko}, \citenamefont
  {Novoselov}, \citenamefont {Roche}, \citenamefont {B{\o}ggild}, \citenamefont
  {Borini}, \citenamefont {Koppens}, \citenamefont {Palermo}, \citenamefont
  {Pugno}, \citenamefont {Garrido}, \citenamefont {Sordan}, \citenamefont
  {Bianco}, \citenamefont {Ballerini}, \citenamefont {Prato}, \citenamefont
  {Lidorikis}, \citenamefont {Kivioja}, \citenamefont {Marinelli},
  \citenamefont {Ryh\"{a}nen}, \citenamefont {Morpurgo}, \citenamefont {Coleman},
  \citenamefont {Nicolosi}, \citenamefont {Colombo}, \citenamefont {Fert},
  \citenamefont {Garcia-Hernandez}, \citenamefont {Bachtold}, \citenamefont
  {Schneider}, \citenamefont {Guinea}, \citenamefont {Dekker}, \citenamefont
  {Barbone}, \citenamefont {Sun}, \citenamefont {Galiotis}, \citenamefont
  {Grigorenko}, \citenamefont {Konstantatos}, \citenamefont {Kis},
  \citenamefont {Katsnelson}, \citenamefont {Vandersypen}, \citenamefont
  {Loiseau}, \citenamefont {Morandi}, \citenamefont {Neumaier}, \citenamefont
  {Treossi}, \citenamefont {Pellegrini}, \citenamefont {Polini}, \citenamefont
  {Tredicucci}, \citenamefont {Williams}, \citenamefont {Hong}, \citenamefont
  {Ahn}, \citenamefont {Kim}, \citenamefont {Zirath}, \citenamefont {van Wees},
  \citenamefont {van~der Zant}, \citenamefont {Occhipinti}, \citenamefont
  {Matteo}, \citenamefont {Kinloch}, \citenamefont {Seyller}, \citenamefont
  {Quesnel}, \citenamefont {Feng}, \citenamefont {Teo}, \citenamefont
  {Rupesinghe}, \citenamefont {Hakonen}, \citenamefont {Neil}, \citenamefont
  {Tannock}, \citenamefont {L\"{o}fwander},\ and\ \citenamefont
  {Kinaret}}]{FER_NS7}%
  \BibitemOpen
  \bibfield  {author} {\bibinfo {author} {\bibfnamefont {A.~C.}\ \bibnamefont
  {Ferrari}}, \bibinfo {author} {\bibfnamefont {F.}~\bibnamefont {Bonaccorso}},
  \bibinfo {author} {\bibfnamefont {V.}~\bibnamefont {Fal'ko}}, \bibinfo
  {author} {\bibfnamefont {K.~S.}\ \bibnamefont {Novoselov}}, \bibinfo {author}
  {\bibfnamefont {S.}~\bibnamefont {Roche}}, \bibinfo {author} {\bibfnamefont
  {P.}~\bibnamefont {B{\o}ggild}}, \bibinfo {author} {\bibfnamefont
  {S.}~\bibnamefont {Borini}}, \bibinfo {author} {\bibfnamefont {F.~H.~L.}\
  \bibnamefont {Koppens}}, \bibinfo {author} {\bibfnamefont {V.}~\bibnamefont
  {Palermo}}, \bibinfo {author} {\bibfnamefont {N.}~\bibnamefont {Pugno}},
  \bibinfo {author} {\bibfnamefont {J.~A.}\ \bibnamefont {Garrido}}, \bibinfo
  {author} {\bibfnamefont {R.}~\bibnamefont {Sordan}}, \bibinfo {author}
  {\bibfnamefont {A.}~\bibnamefont {Bianco}}, \bibinfo {author} {\bibfnamefont
  {L.}~\bibnamefont {Ballerini}}, \bibinfo {author} {\bibfnamefont
  {M.}~\bibnamefont {Prato}}, \bibinfo {author} {\bibfnamefont
  {E.}~\bibnamefont {Lidorikis}}, \bibinfo {author} {\bibfnamefont
  {J.}~\bibnamefont {Kivioja}}, \bibinfo {author} {\bibfnamefont
  {C.}~\bibnamefont {Marinelli}}, \bibinfo {author} {\bibfnamefont
  {T.}~\bibnamefont {Ryh\"{a}nen}}, \bibinfo {author} {\bibfnamefont
  {A.}~\bibnamefont {Morpurgo}}, \bibinfo {author} {\bibfnamefont {J.~N.}\
  \bibnamefont {Coleman}}, \bibinfo {author} {\bibfnamefont {V.}~\bibnamefont
  {Nicolosi}}, \bibinfo {author} {\bibfnamefont {L.}~\bibnamefont {Colombo}},
  \bibinfo {author} {\bibfnamefont {A.}~\bibnamefont {Fert}}, \bibinfo {author}
  {\bibfnamefont {M.}~\bibnamefont {Garcia-Hernandez}}, \bibinfo {author}
  {\bibfnamefont {A.}~\bibnamefont {Bachtold}}, \bibinfo {author}
  {\bibfnamefont {G.~F.}\ \bibnamefont {Schneider}}, \bibinfo {author}
  {\bibfnamefont {F.}~\bibnamefont {Guinea}}, \bibinfo {author} {\bibfnamefont
  {C.}~\bibnamefont {Dekker}}, \bibinfo {author} {\bibfnamefont
  {M.}~\bibnamefont {Barbone}}, \bibinfo {author} {\bibfnamefont
  {Z.}~\bibnamefont {Sun}}, \bibinfo {author} {\bibfnamefont {C.}~\bibnamefont
  {Galiotis}}, \bibinfo {author} {\bibfnamefont {A.~N.}\ \bibnamefont
  {Grigorenko}}, \bibinfo {author} {\bibfnamefont {G.}~\bibnamefont
  {Konstantatos}}, \bibinfo {author} {\bibfnamefont {A.}~\bibnamefont {Kis}},
  \bibinfo {author} {\bibfnamefont {M.}~\bibnamefont {Katsnelson}}, \bibinfo
  {author} {\bibfnamefont {L.}~\bibnamefont {Vandersypen}}, \bibinfo {author}
  {\bibfnamefont {A.}~\bibnamefont {Loiseau}}, \bibinfo {author} {\bibfnamefont
  {V.}~\bibnamefont {Morandi}}, \bibinfo {author} {\bibfnamefont
  {D.}~\bibnamefont {Neumaier}}, \bibinfo {author} {\bibfnamefont
  {E.}~\bibnamefont {Treossi}}, \bibinfo {author} {\bibfnamefont
  {V.}~\bibnamefont {Pellegrini}}, \bibinfo {author} {\bibfnamefont
  {M.}~\bibnamefont {Polini}}, \bibinfo {author} {\bibfnamefont
  {A.}~\bibnamefont {Tredicucci}}, \bibinfo {author} {\bibfnamefont {G.~M.}\
  \bibnamefont {Williams}}, \bibinfo {author} {\bibfnamefont {B.~H.}\
  \bibnamefont {Hong}}, \bibinfo {author} {\bibfnamefont {J.-H.}\ \bibnamefont
  {Ahn}}, \bibinfo {author} {\bibfnamefont {J.~M.}\ \bibnamefont {Kim}},
  \bibinfo {author} {\bibfnamefont {H.}~\bibnamefont {Zirath}}, \bibinfo
  {author} {\bibfnamefont {B.~J.}\ \bibnamefont {van Wees}}, \bibinfo {author}
  {\bibfnamefont {H.}~\bibnamefont {van~der Zant}}, \bibinfo {author}
  {\bibfnamefont {L.}~\bibnamefont {Occhipinti}}, \bibinfo {author}
  {\bibfnamefont {A.~D.}\ \bibnamefont {Matteo}}, \bibinfo {author}
  {\bibfnamefont {I.~A.}\ \bibnamefont {Kinloch}}, \bibinfo {author}
  {\bibfnamefont {T.}~\bibnamefont {Seyller}}, \bibinfo {author} {\bibfnamefont
  {E.}~\bibnamefont {Quesnel}}, \bibinfo {author} {\bibfnamefont
  {X.}~\bibnamefont {Feng}}, \bibinfo {author} {\bibfnamefont {K.}~\bibnamefont
  {Teo}}, \bibinfo {author} {\bibfnamefont {N.}~\bibnamefont {Rupesinghe}},
  \bibinfo {author} {\bibfnamefont {P.}~\bibnamefont {Hakonen}}, \bibinfo
  {author} {\bibfnamefont {S.~R.~T.}\ \bibnamefont {Neil}}, \bibinfo {author}
  {\bibfnamefont {Q.}~\bibnamefont {Tannock}}, \bibinfo {author} {\bibfnamefont
  {T.}~\bibnamefont {L\"{o}fwander}}, \ and\ \bibinfo {author} {\bibfnamefont
  {J.}~\bibnamefont {Kinaret}},\ }\href {\doibase 10.1039/c4nr01600a}
  {\bibfield  {journal} {\bibinfo  {journal} {Nanoscale}\ }\textbf {\bibinfo
  {volume} {7}},\ \bibinfo {pages} {4598} (\bibinfo {year} {2015})}\BibitemShut
  {NoStop}%
\bibitem [{\citenamefont {Lemme}\ \emph {et~al.}(2014)\citenamefont {Lemme},
  \citenamefont {Li}, \citenamefont {Palacios},\ and\ \citenamefont
  {Schwierz}}]{LEM_MRS39}%
  \BibitemOpen
  \bibfield  {author} {\bibinfo {author} {\bibfnamefont {M.~C.}\ \bibnamefont
  {Lemme}}, \bibinfo {author} {\bibfnamefont {L.-J.}\ \bibnamefont {Li}},
  \bibinfo {author} {\bibfnamefont {T.}~\bibnamefont {Palacios}}, \ and\
  \bibinfo {author} {\bibfnamefont {F.}~\bibnamefont {Schwierz}},\ }\href
  {\doibase 10.1557/mrs.2014.138} {\bibfield  {journal} {\bibinfo  {journal}
  {MRS Bull.}\ }\textbf {\bibinfo {volume} {39}},\ \bibinfo {pages} {711}
  (\bibinfo {year} {2014})}\BibitemShut {NoStop}%
\bibitem [{\citenamefont {Bonaccorso}\ \emph {et~al.}(2010)\citenamefont
  {Bonaccorso}, \citenamefont {Sun}, \citenamefont {Hasan},\ and\ \citenamefont
  {Ferrari}}]{BON_NP4}%
  \BibitemOpen
  \bibfield  {author} {\bibinfo {author} {\bibfnamefont {F.}~\bibnamefont
  {Bonaccorso}}, \bibinfo {author} {\bibfnamefont {Z.}~\bibnamefont {Sun}},
  \bibinfo {author} {\bibfnamefont {T.}~\bibnamefont {Hasan}}, \ and\ \bibinfo
  {author} {\bibfnamefont {A.~C.}\ \bibnamefont {Ferrari}},\ }\href {\doibase
  10.1038/nphoton.2010.186} {\bibfield  {journal} {\bibinfo  {journal} {Nat.
  Photonics}\ }\textbf {\bibinfo {volume} {4}},\ \bibinfo {pages} {611}
  (\bibinfo {year} {2010})}\BibitemShut {NoStop}%
\bibitem [{\citenamefont {Han}\ \emph {et~al.}(2014)\citenamefont {Han},
  \citenamefont {Kawakami}, \citenamefont {Gmitra},\ and\ \citenamefont
  {Fabian}}]{HAN_NN9}%
  \BibitemOpen
  \bibfield  {author} {\bibinfo {author} {\bibfnamefont {W.}~\bibnamefont
  {Han}}, \bibinfo {author} {\bibfnamefont {R.~K.}\ \bibnamefont {Kawakami}},
  \bibinfo {author} {\bibfnamefont {M.}~\bibnamefont {Gmitra}}, \ and\ \bibinfo
  {author} {\bibfnamefont {J.}~\bibnamefont {Fabian}},\ }\href {\doibase
  10.1038/nnano.2014.214} {\bibfield  {journal} {\bibinfo  {journal} {Nat.
  Nanotechnol.}\ }\textbf {\bibinfo {volume} {9}},\ \bibinfo {pages} {794}
  (\bibinfo {year} {2014})}\BibitemShut {NoStop}%
\bibitem [{\citenamefont {Poirier}\ and\ \citenamefont
  {Schopfer}(2009)}]{POI_EPJ172}%
  \BibitemOpen
  \bibfield  {author} {\bibinfo {author} {\bibfnamefont {W.}~\bibnamefont
  {Poirier}}\ and\ \bibinfo {author} {\bibfnamefont {F.}~\bibnamefont
  {Schopfer}},\ }\href {\doibase 10.1140/epjst/e2009-01051-5} {\bibfield
  {journal} {\bibinfo  {journal} {The European Physical Journal Special
  Topics}\ }\textbf {\bibinfo {volume} {172}},\ \bibinfo {pages} {207}
  (\bibinfo {year} {2009})}\BibitemShut {NoStop}%
\bibitem [{\citenamefont {Lehman}\ \emph {et~al.}(2011)\citenamefont {Lehman},
  \citenamefont {Lee},\ and\ \citenamefont {Grossman}}]{LEH_AO50}%
  \BibitemOpen
  \bibfield  {author} {\bibinfo {author} {\bibfnamefont {J.~H.}\ \bibnamefont
  {Lehman}}, \bibinfo {author} {\bibfnamefont {B.}~\bibnamefont {Lee}}, \ and\
  \bibinfo {author} {\bibfnamefont {E.~N.}\ \bibnamefont {Grossman}},\ }\href
  {\doibase 10.1364/AO.50.004099} {\bibfield  {journal} {\bibinfo  {journal}
  {Appl. Opt.}\ }\textbf {\bibinfo {volume} {50}},\ \bibinfo {pages} {4099}
  (\bibinfo {year} {2011})}\BibitemShut {NoStop}%
\bibitem [{\citenamefont {Rycerz}\ \emph {et~al.}(2007)\citenamefont {Rycerz},
  \citenamefont {Tworzydlo},\ and\ \citenamefont {Beenakker}}]{RYC_NP3}%
  \BibitemOpen
  \bibfield  {author} {\bibinfo {author} {\bibfnamefont {A.}~\bibnamefont
  {Rycerz}}, \bibinfo {author} {\bibfnamefont {J.}~\bibnamefont {Tworzydlo}}, \
  and\ \bibinfo {author} {\bibfnamefont {C.~W.~J.}\ \bibnamefont {Beenakker}},\
  }\href {\doibase 10.1038/nphys547} {\bibfield  {journal} {\bibinfo  {journal}
  {Nat. Phys.}\ }\textbf {\bibinfo {volume} {3}},\ \bibinfo {pages} {172}
  (\bibinfo {year} {2007})}\BibitemShut {NoStop}%
\bibitem [{\citenamefont {Gorbachev}\ \emph {et~al.}(2014)\citenamefont
  {Gorbachev}, \citenamefont {Song}, \citenamefont {Yu}, \citenamefont
  {Kretinin}, \citenamefont {Withers}, \citenamefont {Cao}, \citenamefont
  {Mishchenko}, \citenamefont {Grigorieva}, \citenamefont {Novoselov},
  \citenamefont {Levitov},\ and\ \citenamefont {Geim}}]{GOR_SCI346}%
  \BibitemOpen
  \bibfield  {author} {\bibinfo {author} {\bibfnamefont {R.~V.}\ \bibnamefont
  {Gorbachev}}, \bibinfo {author} {\bibfnamefont {J.~C.~W.}\ \bibnamefont
  {Song}}, \bibinfo {author} {\bibfnamefont {G.~L.}\ \bibnamefont {Yu}},
  \bibinfo {author} {\bibfnamefont {A.~V.}\ \bibnamefont {Kretinin}}, \bibinfo
  {author} {\bibfnamefont {F.}~\bibnamefont {Withers}}, \bibinfo {author}
  {\bibfnamefont {Y.}~\bibnamefont {Cao}}, \bibinfo {author} {\bibfnamefont
  {A.}~\bibnamefont {Mishchenko}}, \bibinfo {author} {\bibfnamefont {I.~V.}\
  \bibnamefont {Grigorieva}}, \bibinfo {author} {\bibfnamefont {K.~S.}\
  \bibnamefont {Novoselov}}, \bibinfo {author} {\bibfnamefont {L.~S.}\
  \bibnamefont {Levitov}}, \ and\ \bibinfo {author} {\bibfnamefont {A.~K.}\
  \bibnamefont {Geim}},\ }\href {\doibase 10.1126/science.1254966} {\bibfield
  {journal} {\bibinfo  {journal} {Science}\ }\textbf {\bibinfo {volume}
  {346}},\ \bibinfo {pages} {448} (\bibinfo {year} {2014})}\BibitemShut
  {NoStop}%
\bibitem [{\citenamefont {Ribeiro-Palau}\ \emph {et~al.}(2015)\citenamefont
  {Ribeiro-Palau}, \citenamefont {Lafont}, \citenamefont {Brun-Picard},
  \citenamefont {Kazazis}, \citenamefont {Michon}, \citenamefont {Cheynis},
  \citenamefont {Couturaud}, \citenamefont {Consejo}, \citenamefont {Jouault},
  \citenamefont {Poirier},\ and\ \citenamefont {Schopfer}}]{RIB_NN10}%
  \BibitemOpen
  \bibfield  {author} {\bibinfo {author} {\bibfnamefont {R.}~\bibnamefont
  {Ribeiro-Palau}}, \bibinfo {author} {\bibfnamefont {F.}~\bibnamefont
  {Lafont}}, \bibinfo {author} {\bibfnamefont {J.}~\bibnamefont {Brun-Picard}},
  \bibinfo {author} {\bibfnamefont {D.}~\bibnamefont {Kazazis}}, \bibinfo
  {author} {\bibfnamefont {A.}~\bibnamefont {Michon}}, \bibinfo {author}
  {\bibfnamefont {F.}~\bibnamefont {Cheynis}}, \bibinfo {author} {\bibfnamefont
  {O.}~\bibnamefont {Couturaud}}, \bibinfo {author} {\bibfnamefont
  {C.}~\bibnamefont {Consejo}}, \bibinfo {author} {\bibfnamefont
  {B.}~\bibnamefont {Jouault}}, \bibinfo {author} {\bibfnamefont
  {W.}~\bibnamefont {Poirier}}, \ and\ \bibinfo {author} {\bibfnamefont
  {F.}~\bibnamefont {Schopfer}},\ }\href {\doibase 10.1038/nnano.2015.192}
  {\bibfield  {journal} {\bibinfo  {journal} {Nat. Nanotechnol.}\ }\textbf
  {\bibinfo {volume} {10}},\ \bibinfo {pages} {965} (\bibinfo {year}
  {2015})}\BibitemShut {NoStop}%
\bibitem [{\citenamefont {Wu}\ \emph {et~al.}(2012)\citenamefont {Wu},
  \citenamefont {Jenkins}, \citenamefont {Valdes-Garcia}, \citenamefont
  {Farmer}, \citenamefont {Zhu}, \citenamefont {Bol}, \citenamefont
  {Dimitrakopoulos}, \citenamefont {Zhu}, \citenamefont {Xia}, \citenamefont
  {Avouris},\ and\ \citenamefont {Lin}}]{WU_NL12}%
  \BibitemOpen
  \bibfield  {author} {\bibinfo {author} {\bibfnamefont {Y.}~\bibnamefont
  {Wu}}, \bibinfo {author} {\bibfnamefont {K.~A.}\ \bibnamefont {Jenkins}},
  \bibinfo {author} {\bibfnamefont {A.}~\bibnamefont {Valdes-Garcia}}, \bibinfo
  {author} {\bibfnamefont {D.~B.}\ \bibnamefont {Farmer}}, \bibinfo {author}
  {\bibfnamefont {Y.}~\bibnamefont {Zhu}}, \bibinfo {author} {\bibfnamefont
  {A.~A.}\ \bibnamefont {Bol}}, \bibinfo {author} {\bibfnamefont
  {C.}~\bibnamefont {Dimitrakopoulos}}, \bibinfo {author} {\bibfnamefont
  {W.}~\bibnamefont {Zhu}}, \bibinfo {author} {\bibfnamefont {F.}~\bibnamefont
  {Xia}}, \bibinfo {author} {\bibfnamefont {P.}~\bibnamefont {Avouris}}, \ and\
  \bibinfo {author} {\bibfnamefont {Y.-M.}\ \bibnamefont {Lin}},\ }\href
  {\doibase 10.1021/nl300904k} {\bibfield  {journal} {\bibinfo  {journal} {Nano
  Lett.}\ }\textbf {\bibinfo {volume} {12}},\ \bibinfo {pages} {3062} (\bibinfo
  {year} {2012})}\BibitemShut {NoStop}%
\bibitem [{\citenamefont {Guo}\ \emph {et~al.}(2013)\citenamefont {Guo},
  \citenamefont {Dong}, \citenamefont {Chakraborty}, \citenamefont {Lourenco},
  \citenamefont {Palmer}, \citenamefont {Hu}, \citenamefont {Ruan},
  \citenamefont {Hankinson}, \citenamefont {Kunc}, \citenamefont {Cressler},
  \citenamefont {Berger},\ and\ \citenamefont {de~Heer}}]{GUO_NL13}%
  \BibitemOpen
  \bibfield  {author} {\bibinfo {author} {\bibfnamefont {Z.}~\bibnamefont
  {Guo}}, \bibinfo {author} {\bibfnamefont {R.}~\bibnamefont {Dong}}, \bibinfo
  {author} {\bibfnamefont {P.~S.}\ \bibnamefont {Chakraborty}}, \bibinfo
  {author} {\bibfnamefont {N.}~\bibnamefont {Lourenco}}, \bibinfo {author}
  {\bibfnamefont {J.}~\bibnamefont {Palmer}}, \bibinfo {author} {\bibfnamefont
  {Y.}~\bibnamefont {Hu}}, \bibinfo {author} {\bibfnamefont {M.}~\bibnamefont
  {Ruan}}, \bibinfo {author} {\bibfnamefont {J.}~\bibnamefont {Hankinson}},
  \bibinfo {author} {\bibfnamefont {J.}~\bibnamefont {Kunc}}, \bibinfo {author}
  {\bibfnamefont {J.~D.}\ \bibnamefont {Cressler}}, \bibinfo {author}
  {\bibfnamefont {C.}~\bibnamefont {Berger}}, \ and\ \bibinfo {author}
  {\bibfnamefont {W.~A.}\ \bibnamefont {de~Heer}},\ }\href {\doibase
  10.1021/nl303587r} {\bibfield  {journal} {\bibinfo  {journal} {Nano Lett.}\
  }\textbf {\bibinfo {volume} {13}},\ \bibinfo {pages} {942} (\bibinfo {year}
  {2013})}\BibitemShut {NoStop}%
\bibitem [{\citenamefont {Withers}\ \emph {et~al.}(2015)\citenamefont
  {Withers}, \citenamefont {Pozo-Zamudio}, \citenamefont {Mishchenko},
  \citenamefont {Rooney}, \citenamefont {Gholinia}, \citenamefont {Watanabe},
  \citenamefont {Taniguchi}, \citenamefont {Haigh}, \citenamefont {Geim},
  \citenamefont {Tartakovskii},\ and\ \citenamefont {Novoselov}}]{WIT_NM14}%
  \BibitemOpen
  \bibfield  {author} {\bibinfo {author} {\bibfnamefont {F.}~\bibnamefont
  {Withers}}, \bibinfo {author} {\bibfnamefont {O.~D.}\ \bibnamefont
  {Pozo-Zamudio}}, \bibinfo {author} {\bibfnamefont {A.}~\bibnamefont
  {Mishchenko}}, \bibinfo {author} {\bibfnamefont {A.~P.}\ \bibnamefont
  {Rooney}}, \bibinfo {author} {\bibfnamefont {A.}~\bibnamefont {Gholinia}},
  \bibinfo {author} {\bibfnamefont {K.}~\bibnamefont {Watanabe}}, \bibinfo
  {author} {\bibfnamefont {T.}~\bibnamefont {Taniguchi}}, \bibinfo {author}
  {\bibfnamefont {S.~J.}\ \bibnamefont {Haigh}}, \bibinfo {author}
  {\bibfnamefont {A.~K.}\ \bibnamefont {Geim}}, \bibinfo {author}
  {\bibfnamefont {A.~I.}\ \bibnamefont {Tartakovskii}}, \ and\ \bibinfo
  {author} {\bibfnamefont {K.~S.}\ \bibnamefont {Novoselov}},\ }\href {\doibase
  10.1038/nmat4205} {\bibfield  {journal} {\bibinfo  {journal} {Nat. Mater.}\
  }\textbf {\bibinfo {volume} {14}},\ \bibinfo {pages} {301} (\bibinfo {year}
  {2015})}\BibitemShut {NoStop}%
\bibitem [{\citenamefont {Son}\ \emph {et~al.}(2006)\citenamefont {Son},
  \citenamefont {Cohen},\ and\ \citenamefont {Louie}}]{SON_PRL97}%
  \BibitemOpen
  \bibfield  {author} {\bibinfo {author} {\bibfnamefont {Y.-W.}\ \bibnamefont
  {Son}}, \bibinfo {author} {\bibfnamefont {M.~L.}\ \bibnamefont {Cohen}}, \
  and\ \bibinfo {author} {\bibfnamefont {S.~G.}\ \bibnamefont {Louie}},\ }\href
  {\doibase 10.1103/physrevlett.97.216803} {\bibfield  {journal} {\bibinfo
  {journal} {Phys. Rev. Lett.}\ }\textbf {\bibinfo {volume} {97}},\ \bibinfo
  {pages} {216803} (\bibinfo {year} {2006})}\BibitemShut {NoStop}%
\bibitem [{\citenamefont {Biel}\ \emph {et~al.}(2009)\citenamefont {Biel},
  \citenamefont {Triozon}, \citenamefont {Blase},\ and\ \citenamefont
  {Roche}}]{BIE_NL9}%
  \BibitemOpen
  \bibfield  {author} {\bibinfo {author} {\bibfnamefont {B.}~\bibnamefont
  {Biel}}, \bibinfo {author} {\bibfnamefont {F.}~\bibnamefont {Triozon}},
  \bibinfo {author} {\bibfnamefont {X.}~\bibnamefont {Blase}}, \ and\ \bibinfo
  {author} {\bibfnamefont {S.}~\bibnamefont {Roche}},\ }\href {\doibase
  10.1021/nl901226s} {\bibfield  {journal} {\bibinfo  {journal} {Nano Lett.}\
  }\textbf {\bibinfo {volume} {9}},\ \bibinfo {pages} {2725} (\bibinfo {year}
  {2009})}\BibitemShut {NoStop}%
\bibitem [{\citenamefont {Marconcini}\ \emph {et~al.}(2012)\citenamefont
  {Marconcini}, \citenamefont {Cresti}, \citenamefont {Triozon}, \citenamefont
  {Fiori}, \citenamefont {Biel}, \citenamefont {Niquet}, \citenamefont
  {Macucci},\ and\ \citenamefont {Roche}}]{MAR_ACSN6}%
  \BibitemOpen
  \bibfield  {author} {\bibinfo {author} {\bibfnamefont {P.}~\bibnamefont
  {Marconcini}}, \bibinfo {author} {\bibfnamefont {A.}~\bibnamefont {Cresti}},
  \bibinfo {author} {\bibfnamefont {F.}~\bibnamefont {Triozon}}, \bibinfo
  {author} {\bibfnamefont {G.}~\bibnamefont {Fiori}}, \bibinfo {author}
  {\bibfnamefont {B.}~\bibnamefont {Biel}}, \bibinfo {author} {\bibfnamefont
  {Y.-M.}\ \bibnamefont {Niquet}}, \bibinfo {author} {\bibfnamefont
  {M.}~\bibnamefont {Macucci}}, \ and\ \bibinfo {author} {\bibfnamefont
  {S.}~\bibnamefont {Roche}},\ }\href {\doibase 10.1021/nn3024046} {\bibfield
  {journal} {\bibinfo  {journal} {{ACS} Nano}\ }\textbf {\bibinfo {volume}
  {6}},\ \bibinfo {pages} {7942} (\bibinfo {year} {2012})}\BibitemShut
  {NoStop}%
\bibitem [{\citenamefont {Castro}\ \emph {et~al.}(2007)\citenamefont {Castro},
  \citenamefont {Novoselov}, \citenamefont {Morozov}, \citenamefont {Peres},
  \citenamefont {dos Santos}, \citenamefont {Nilsson}, \citenamefont {Guinea},
  \citenamefont {Geim},\ and\ \citenamefont {Neto}}]{CAS_PRL99}%
  \BibitemOpen
  \bibfield  {author} {\bibinfo {author} {\bibfnamefont {E.~V.}\ \bibnamefont
  {Castro}}, \bibinfo {author} {\bibfnamefont {K.~S.}\ \bibnamefont
  {Novoselov}}, \bibinfo {author} {\bibfnamefont {S.~V.}\ \bibnamefont
  {Morozov}}, \bibinfo {author} {\bibfnamefont {N.~M.~R.}\ \bibnamefont
  {Peres}}, \bibinfo {author} {\bibfnamefont {J.~M. B.~L.}\ \bibnamefont {dos
  Santos}}, \bibinfo {author} {\bibfnamefont {J.}~\bibnamefont {Nilsson}},
  \bibinfo {author} {\bibfnamefont {F.}~\bibnamefont {Guinea}}, \bibinfo
  {author} {\bibfnamefont {A.~K.}\ \bibnamefont {Geim}}, \ and\ \bibinfo
  {author} {\bibfnamefont {A.~H.~C.}\ \bibnamefont {Neto}},\ }\href {\doibase
  10.1103/physrevlett.99.216802} {\bibfield  {journal} {\bibinfo  {journal}
  {Phys. Rev. Lett.}\ }\textbf {\bibinfo {volume} {99}},\ \bibinfo {pages}
  {216802} (\bibinfo {year} {2007})}\BibitemShut {NoStop}%
\bibitem [{\citenamefont {Oostinga}\ \emph {et~al.}(2008)\citenamefont
  {Oostinga}, \citenamefont {Heersche}, \citenamefont {Liu}, \citenamefont
  {Morpurgo},\ and\ \citenamefont {Vandersypen}}]{OOS_NM7}%
  \BibitemOpen
  \bibfield  {author} {\bibinfo {author} {\bibfnamefont {J.~B.}\ \bibnamefont
  {Oostinga}}, \bibinfo {author} {\bibfnamefont {H.~B.}\ \bibnamefont
  {Heersche}}, \bibinfo {author} {\bibfnamefont {X.}~\bibnamefont {Liu}},
  \bibinfo {author} {\bibfnamefont {A.~F.}\ \bibnamefont {Morpurgo}}, \ and\
  \bibinfo {author} {\bibfnamefont {L.~M.~K.}\ \bibnamefont {Vandersypen}},\
  }\href {\doibase 10.1038/nmat2082} {\bibfield  {journal} {\bibinfo  {journal}
  {Nat. Mater.}\ }\textbf {\bibinfo {volume} {7}},\ \bibinfo {pages} {151}
  (\bibinfo {year} {2008})}\BibitemShut {NoStop}%
\bibitem [{\citenamefont {Mu{\~{n}}oz}\ and\ \citenamefont
  {G{\'{o}}mez-Aleixandre}(2013)}]{MUN_CVD19}%
  \BibitemOpen
  \bibfield  {author} {\bibinfo {author} {\bibfnamefont {R.}~\bibnamefont
  {Mu{\~{n}}oz}}\ and\ \bibinfo {author} {\bibfnamefont {C.}~\bibnamefont
  {G{\'{o}}mez-Aleixandre}},\ }\href {\doibase 10.1002/cvde.201300051}
  {\bibfield  {journal} {\bibinfo  {journal} {Chem. Vap. Deposition}\ }\textbf
  {\bibinfo {volume} {19}},\ \bibinfo {pages} {297} (\bibinfo {year}
  {2013})}\BibitemShut {NoStop}%
\bibitem [{\citenamefont {Huang}\ \emph {et~al.}(2011)\citenamefont {Huang},
  \citenamefont {Ruiz-Vargas}, \citenamefont {van~der Zande}, \citenamefont
  {Whitney}, \citenamefont {Levendorf}, \citenamefont {Kevek}, \citenamefont
  {Garg}, \citenamefont {Alden}, \citenamefont {Hustedt}, \citenamefont {Zhu},
  \citenamefont {Park}, \citenamefont {McEuen},\ and\ \citenamefont
  {Muller}}]{HUA_NAT469}%
  \BibitemOpen
  \bibfield  {author} {\bibinfo {author} {\bibfnamefont {P.~Y.}\ \bibnamefont
  {Huang}}, \bibinfo {author} {\bibfnamefont {C.~S.}\ \bibnamefont
  {Ruiz-Vargas}}, \bibinfo {author} {\bibfnamefont {A.~M.}\ \bibnamefont
  {van~der Zande}}, \bibinfo {author} {\bibfnamefont {W.~S.}\ \bibnamefont
  {Whitney}}, \bibinfo {author} {\bibfnamefont {M.~P.}\ \bibnamefont
  {Levendorf}}, \bibinfo {author} {\bibfnamefont {J.~W.}\ \bibnamefont
  {Kevek}}, \bibinfo {author} {\bibfnamefont {S.}~\bibnamefont {Garg}},
  \bibinfo {author} {\bibfnamefont {J.~S.}\ \bibnamefont {Alden}}, \bibinfo
  {author} {\bibfnamefont {C.~J.}\ \bibnamefont {Hustedt}}, \bibinfo {author}
  {\bibfnamefont {Y.}~\bibnamefont {Zhu}}, \bibinfo {author} {\bibfnamefont
  {J.}~\bibnamefont {Park}}, \bibinfo {author} {\bibfnamefont {P.~L.}\
  \bibnamefont {McEuen}}, \ and\ \bibinfo {author} {\bibfnamefont {D.~A.}\
  \bibnamefont {Muller}},\ }\href {\doibase 10.1038/nature09718} {\bibfield
  {journal} {\bibinfo  {journal} {Nature}\ }\textbf {\bibinfo {volume} {469}},\
  \bibinfo {pages} {389} (\bibinfo {year} {2011})}\BibitemShut {NoStop}%
\bibitem [{\citenamefont {Cockayne}\ \emph {et~al.}(2011)\citenamefont
  {Cockayne}, \citenamefont {Rutter}, \citenamefont {Guisinger}, \citenamefont
  {Crain}, \citenamefont {First},\ and\ \citenamefont {Stroscio}}]{COC_PRB83}%
  \BibitemOpen
  \bibfield  {author} {\bibinfo {author} {\bibfnamefont {E.}~\bibnamefont
  {Cockayne}}, \bibinfo {author} {\bibfnamefont {G.~M.}\ \bibnamefont
  {Rutter}}, \bibinfo {author} {\bibfnamefont {N.~P.}\ \bibnamefont
  {Guisinger}}, \bibinfo {author} {\bibfnamefont {J.~N.}\ \bibnamefont
  {Crain}}, \bibinfo {author} {\bibfnamefont {P.~N.}\ \bibnamefont {First}}, \
  and\ \bibinfo {author} {\bibfnamefont {J.~A.}\ \bibnamefont {Stroscio}},\
  }\href {\doibase 10.1103/physrevb.83.195425} {\bibfield  {journal} {\bibinfo
  {journal} {Phys. Rev. B}\ }\textbf {\bibinfo {volume} {83}},\ \bibinfo
  {pages} {195425} (\bibinfo {year} {2011})}\BibitemShut {NoStop}%
\bibitem [{\citenamefont {Batzill}(2012)}]{BAT_SSR67}%
  \BibitemOpen
  \bibfield  {author} {\bibinfo {author} {\bibfnamefont {M.}~\bibnamefont
  {Batzill}},\ }\href {\doibase 10.1016/j.surfrep.2011.12.001} {\bibfield
  {journal} {\bibinfo  {journal} {Surf. Sci. Rep.}\ }\textbf {\bibinfo {volume}
  {67}},\ \bibinfo {pages} {83} (\bibinfo {year} {2012})}\BibitemShut {NoStop}%
\bibitem [{\citenamefont {Cockayne}(2012)}]{COC_PRB85}%
  \BibitemOpen
  \bibfield  {author} {\bibinfo {author} {\bibfnamefont {E.}~\bibnamefont
  {Cockayne}},\ }\href {\doibase 10.1103/physrevb.85.125409} {\bibfield
  {journal} {\bibinfo  {journal} {Phys. Rev. B}\ }\textbf {\bibinfo {volume}
  {85}},\ \bibinfo {pages} {125409} (\bibinfo {year} {2012})}\BibitemShut
  {NoStop}%
\bibitem [{\citenamefont {Hersam}(2015)}]{HER_PCL6}%
  \BibitemOpen
  \bibfield  {author} {\bibinfo {author} {\bibfnamefont {M.~C.}\ \bibnamefont
  {Hersam}},\ }\href {\doibase 10.1021/acs.jpclett.5b01218} {\bibfield
  {journal} {\bibinfo  {journal} {The Journal of Physical Chemistry Letters}\
  }\textbf {\bibinfo {volume} {6}},\ \bibinfo {pages} {2738} (\bibinfo {year}
  {2015})}\BibitemShut {NoStop}%
\bibitem [{\citenamefont {Bir{\'{o}}}\ and\ \citenamefont
  {Lambin}(2013)}]{BIR_NJP15}%
  \BibitemOpen
  \bibfield  {author} {\bibinfo {author} {\bibfnamefont {L.~P.}\ \bibnamefont
  {Bir{\'{o}}}}\ and\ \bibinfo {author} {\bibfnamefont {P.}~\bibnamefont
  {Lambin}},\ }\href {\doibase 10.1088/1367-2630/15/3/035024} {\bibfield
  {journal} {\bibinfo  {journal} {New J. Phys.}\ }\textbf {\bibinfo {volume}
  {15}},\ \bibinfo {pages} {035024} (\bibinfo {year} {2013})}\BibitemShut
  {NoStop}%
\bibitem [{\citenamefont {Terrones}\ \emph {et~al.}(2012)\citenamefont
  {Terrones}, \citenamefont {Lv}, \citenamefont {Terrones},\ and\ \citenamefont
  {Dresselhaus}}]{TER_RPP75}%
  \BibitemOpen
  \bibfield  {author} {\bibinfo {author} {\bibfnamefont {H.}~\bibnamefont
  {Terrones}}, \bibinfo {author} {\bibfnamefont {R.}~\bibnamefont {Lv}},
  \bibinfo {author} {\bibfnamefont {M.}~\bibnamefont {Terrones}}, \ and\
  \bibinfo {author} {\bibfnamefont {M.~S.}\ \bibnamefont {Dresselhaus}},\
  }\href {\doibase 10.1088/0034-4885/75/6/062501} {\bibfield  {journal}
  {\bibinfo  {journal} {Rep. Prog. Phys.}\ }\textbf {\bibinfo {volume} {75}},\
  \bibinfo {pages} {062501} (\bibinfo {year} {2012})}\BibitemShut {NoStop}%
\bibitem [{\citenamefont {Banhart}\ \emph {et~al.}(2011)\citenamefont
  {Banhart}, \citenamefont {Kotakoski},\ and\ \citenamefont
  {Krasheninnikov}}]{BAN_ACSN5}%
  \BibitemOpen
  \bibfield  {author} {\bibinfo {author} {\bibfnamefont {F.}~\bibnamefont
  {Banhart}}, \bibinfo {author} {\bibfnamefont {J.}~\bibnamefont {Kotakoski}},
  \ and\ \bibinfo {author} {\bibfnamefont {A.~V.}\ \bibnamefont
  {Krasheninnikov}},\ }\href {\doibase 10.1021/nn102598m} {\bibfield  {journal}
  {\bibinfo  {journal} {{ACS} Nano}\ }\textbf {\bibinfo {volume} {5}},\
  \bibinfo {pages} {26} (\bibinfo {year} {2011})}\BibitemShut {NoStop}%
\bibitem [{\citenamefont {Lu}\ \emph {et~al.}(2013)\citenamefont {Lu},
  \citenamefont {Bao}, \citenamefont {Su},\ and\ \citenamefont
  {Loh}}]{LU_ACSN7}%
  \BibitemOpen
  \bibfield  {author} {\bibinfo {author} {\bibfnamefont {J.}~\bibnamefont
  {Lu}}, \bibinfo {author} {\bibfnamefont {Y.}~\bibnamefont {Bao}}, \bibinfo
  {author} {\bibfnamefont {C.~L.}\ \bibnamefont {Su}}, \ and\ \bibinfo {author}
  {\bibfnamefont {K.~P.}\ \bibnamefont {Loh}},\ }\href {\doibase
  10.1021/nn4051248} {\bibfield  {journal} {\bibinfo  {journal} {{ACS} Nano}\
  }\textbf {\bibinfo {volume} {7}},\ \bibinfo {pages} {8350} (\bibinfo {year}
  {2013})}\BibitemShut {NoStop}%
\bibitem [{\citenamefont {Yazyev}\ and\ \citenamefont {Louie}(2010)}]{YAZ_NM9}%
  \BibitemOpen
  \bibfield  {author} {\bibinfo {author} {\bibfnamefont {O.~V.}\ \bibnamefont
  {Yazyev}}\ and\ \bibinfo {author} {\bibfnamefont {S.~G.}\ \bibnamefont
  {Louie}},\ }\href {\doibase 10.1038/nmat2830} {\bibfield  {journal} {\bibinfo
   {journal} {Nat. Mater.}\ }\textbf {\bibinfo {volume} {9}},\ \bibinfo {pages}
  {806} (\bibinfo {year} {2010})}\BibitemShut {NoStop}%
\bibitem [{\citenamefont {Lherbier}\ \emph {et~al.}(2012)\citenamefont
  {Lherbier}, \citenamefont {Dubois}, \citenamefont {Declerck}, \citenamefont
  {Niquet}, \citenamefont {Roche},\ and\ \citenamefont {Charlier}}]{LHE_PRB86}%
  \BibitemOpen
  \bibfield  {author} {\bibinfo {author} {\bibfnamefont {A.}~\bibnamefont
  {Lherbier}}, \bibinfo {author} {\bibfnamefont {S.~M.-M.}\ \bibnamefont
  {Dubois}}, \bibinfo {author} {\bibfnamefont {X.}~\bibnamefont {Declerck}},
  \bibinfo {author} {\bibfnamefont {Y.-M.}\ \bibnamefont {Niquet}}, \bibinfo
  {author} {\bibfnamefont {S.}~\bibnamefont {Roche}}, \ and\ \bibinfo {author}
  {\bibfnamefont {J.-C.}\ \bibnamefont {Charlier}},\ }\href {\doibase
  10.1103/physrevb.86.075402} {\bibfield  {journal} {\bibinfo  {journal} {Phys.
  Rev. B}\ }\textbf {\bibinfo {volume} {86}},\ \bibinfo {pages} {075402}
  (\bibinfo {year} {2012})}\BibitemShut {NoStop}%
\bibitem [{\citenamefont {Tuan}\ \emph {et~al.}(2013)\citenamefont {Tuan},
  \citenamefont {Kotakoski}, \citenamefont {Louvet}, \citenamefont {Ortmann},
  \citenamefont {Meyer},\ and\ \citenamefont {Roche}}]{VAN_NL13}%
  \BibitemOpen
  \bibfield  {author} {\bibinfo {author} {\bibfnamefont {D.~V.}\ \bibnamefont
  {Tuan}}, \bibinfo {author} {\bibfnamefont {J.}~\bibnamefont {Kotakoski}},
  \bibinfo {author} {\bibfnamefont {T.}~\bibnamefont {Louvet}}, \bibinfo
  {author} {\bibfnamefont {F.}~\bibnamefont {Ortmann}}, \bibinfo {author}
  {\bibfnamefont {J.~C.}\ \bibnamefont {Meyer}}, \ and\ \bibinfo {author}
  {\bibfnamefont {S.}~\bibnamefont {Roche}},\ }\href {\doibase
  10.1021/nl400321r} {\bibfield  {journal} {\bibinfo  {journal} {Nano Lett.}\
  }\textbf {\bibinfo {volume} {13}},\ \bibinfo {pages} {1730} (\bibinfo {year}
  {2013})}\BibitemShut {NoStop}%
\bibitem [{\citenamefont {Lafont}\ \emph {et~al.}(2014)\citenamefont {Lafont},
  \citenamefont {Ribeiro-Palau}, \citenamefont {Han}, \citenamefont {Cresti},
  \citenamefont {Delvall{\'{e}}e}, \citenamefont {Cummings}, \citenamefont
  {Roche}, \citenamefont {Bouchiat}, \citenamefont {Ducourtieux}, \citenamefont
  {Schopfer},\ and\ \citenamefont {Poirier}}]{LAF_PRB90}%
  \BibitemOpen
  \bibfield  {author} {\bibinfo {author} {\bibfnamefont {F.}~\bibnamefont
  {Lafont}}, \bibinfo {author} {\bibfnamefont {R.}~\bibnamefont
  {Ribeiro-Palau}}, \bibinfo {author} {\bibfnamefont {Z.}~\bibnamefont {Han}},
  \bibinfo {author} {\bibfnamefont {A.}~\bibnamefont {Cresti}}, \bibinfo
  {author} {\bibfnamefont {A.}~\bibnamefont {Delvall{\'{e}}e}}, \bibinfo
  {author} {\bibfnamefont {A.~W.}\ \bibnamefont {Cummings}}, \bibinfo {author}
  {\bibfnamefont {S.}~\bibnamefont {Roche}}, \bibinfo {author} {\bibfnamefont
  {V.}~\bibnamefont {Bouchiat}}, \bibinfo {author} {\bibfnamefont
  {S.}~\bibnamefont {Ducourtieux}}, \bibinfo {author} {\bibfnamefont
  {F.}~\bibnamefont {Schopfer}}, \ and\ \bibinfo {author} {\bibfnamefont
  {W.}~\bibnamefont {Poirier}},\ }\href {\doibase 10.1103/physrevb.90.115422}
  {\bibfield  {journal} {\bibinfo  {journal} {Phys. Rev. B}\ }\textbf {\bibinfo
  {volume} {90}},\ \bibinfo {pages} {115422} (\bibinfo {year}
  {2014})}\BibitemShut {NoStop}%
\bibitem [{\citenamefont {Vancs{\'{o}}}\ \emph {et~al.}(2013)\citenamefont
  {Vancs{\'{o}}}, \citenamefont {M{\'{a}}rk}, \citenamefont {Lambin},
  \citenamefont {Mayer}, \citenamefont {Kim}, \citenamefont {Hwang},\ and\
  \citenamefont {Bir{\'{o}}}}]{Vancso2013}%
  \BibitemOpen
  \bibfield  {author} {\bibinfo {author} {\bibfnamefont {P.}~\bibnamefont
  {Vancs{\'{o}}}}, \bibinfo {author} {\bibfnamefont {G.~I.}\ \bibnamefont
  {M{\'{a}}rk}}, \bibinfo {author} {\bibfnamefont {P.}~\bibnamefont {Lambin}},
  \bibinfo {author} {\bibfnamefont {A.}~\bibnamefont {Mayer}}, \bibinfo
  {author} {\bibfnamefont {Y.-S.}\ \bibnamefont {Kim}}, \bibinfo {author}
  {\bibfnamefont {C.}~\bibnamefont {Hwang}}, \ and\ \bibinfo {author}
  {\bibfnamefont {L.~P.}\ \bibnamefont {Bir{\'{o}}}},\ }\href {\doibase
  10.1016/j.carbon.2013.07.041} {\bibfield  {journal} {\bibinfo  {journal}
  {Carbon}\ }\textbf {\bibinfo {volume} {64}},\ \bibinfo {pages} {101}
  (\bibinfo {year} {2013})}\BibitemShut {NoStop}%
\bibitem [{\citenamefont {Tsen}\ \emph {et~al.}(2012)\citenamefont {Tsen},
  \citenamefont {Brown}, \citenamefont {Levendorf}, \citenamefont {Ghahari},
  \citenamefont {Huang}, \citenamefont {Havener}, \citenamefont {Ruiz-Vargas},
  \citenamefont {Muller}, \citenamefont {Kim},\ and\ \citenamefont
  {Park}}]{TSE_SCI336}%
  \BibitemOpen
  \bibfield  {author} {\bibinfo {author} {\bibfnamefont {A.~W.}\ \bibnamefont
  {Tsen}}, \bibinfo {author} {\bibfnamefont {L.}~\bibnamefont {Brown}},
  \bibinfo {author} {\bibfnamefont {M.~P.}\ \bibnamefont {Levendorf}}, \bibinfo
  {author} {\bibfnamefont {F.}~\bibnamefont {Ghahari}}, \bibinfo {author}
  {\bibfnamefont {P.~Y.}\ \bibnamefont {Huang}}, \bibinfo {author}
  {\bibfnamefont {R.~W.}\ \bibnamefont {Havener}}, \bibinfo {author}
  {\bibfnamefont {C.~S.}\ \bibnamefont {Ruiz-Vargas}}, \bibinfo {author}
  {\bibfnamefont {D.~A.}\ \bibnamefont {Muller}}, \bibinfo {author}
  {\bibfnamefont {P.}~\bibnamefont {Kim}}, \ and\ \bibinfo {author}
  {\bibfnamefont {J.}~\bibnamefont {Park}},\ }\href {\doibase
  10.1126/science.1218948} {\bibfield  {journal} {\bibinfo  {journal}
  {Science}\ }\textbf {\bibinfo {volume} {336}},\ \bibinfo {pages} {1143}
  (\bibinfo {year} {2012})}\BibitemShut {NoStop}%
\bibitem [{\citenamefont {Kurasch}\ \emph {et~al.}(2012)\citenamefont
  {Kurasch}, \citenamefont {Kotakoski}, \citenamefont {Lehtinen}, \citenamefont
  {Sk{\'{a}}kalov{\'{a}}}, \citenamefont {Smet}, \citenamefont {Krill},
  \citenamefont {Krasheninnikov},\ and\ \citenamefont {Kaiser}}]{KUR_NL12}%
  \BibitemOpen
  \bibfield  {author} {\bibinfo {author} {\bibfnamefont {S.}~\bibnamefont
  {Kurasch}}, \bibinfo {author} {\bibfnamefont {J.}~\bibnamefont {Kotakoski}},
  \bibinfo {author} {\bibfnamefont {O.}~\bibnamefont {Lehtinen}}, \bibinfo
  {author} {\bibfnamefont {V.}~\bibnamefont {Sk{\'{a}}kalov{\'{a}}}}, \bibinfo
  {author} {\bibfnamefont {J.}~\bibnamefont {Smet}}, \bibinfo {author}
  {\bibfnamefont {C.~E.}\ \bibnamefont {Krill}}, \bibinfo {author}
  {\bibfnamefont {A.~V.}\ \bibnamefont {Krasheninnikov}}, \ and\ \bibinfo
  {author} {\bibfnamefont {U.}~\bibnamefont {Kaiser}},\ }\href {\doibase
  10.1021/nl301141g} {\bibfield  {journal} {\bibinfo  {journal} {Nano Lett.}\
  }\textbf {\bibinfo {volume} {12}},\ \bibinfo {pages} {3168} (\bibinfo {year}
  {2012})}\BibitemShut {NoStop}%
\bibitem [{\citenamefont {Lehtinen}\ \emph {et~al.}(2013)\citenamefont
  {Lehtinen}, \citenamefont {Kurasch}, \citenamefont {Krasheninnikov},\ and\
  \citenamefont {Kaiser}}]{LEH_NC4}%
  \BibitemOpen
  \bibfield  {author} {\bibinfo {author} {\bibfnamefont {O.}~\bibnamefont
  {Lehtinen}}, \bibinfo {author} {\bibfnamefont {S.}~\bibnamefont {Kurasch}},
  \bibinfo {author} {\bibfnamefont {A.}~\bibnamefont {Krasheninnikov}}, \ and\
  \bibinfo {author} {\bibfnamefont {U.}~\bibnamefont {Kaiser}},\ }\href
  {\doibase 10.1038/ncomms3098} {\bibfield  {journal} {\bibinfo  {journal}
  {Nat. Commun.}\ }\textbf {\bibinfo {volume} {4}},\ \bibinfo {pages} {2098}
  (\bibinfo {year} {2013})}\BibitemShut {NoStop}%
\bibitem [{\citenamefont {Lee}\ \emph {et~al.}(2013)\citenamefont {Lee},
  \citenamefont {Yoon}, \citenamefont {Wang},\ and\ \citenamefont
  {Ho}}]{LEE_JPCM25}%
  \BibitemOpen
  \bibfield  {author} {\bibinfo {author} {\bibfnamefont {G.-D.}\ \bibnamefont
  {Lee}}, \bibinfo {author} {\bibfnamefont {E.}~\bibnamefont {Yoon}}, \bibinfo
  {author} {\bibfnamefont {C.-Z.}\ \bibnamefont {Wang}}, \ and\ \bibinfo
  {author} {\bibfnamefont {K.-M.}\ \bibnamefont {Ho}},\ }\href {\doibase
  10.1088/0953-8984/25/15/155301} {\bibfield  {journal} {\bibinfo  {journal}
  {J. Phys.: Condens. Matter}\ }\textbf {\bibinfo {volume} {25}},\ \bibinfo
  {pages} {155301} (\bibinfo {year} {2013})}\BibitemShut {NoStop}%
\bibitem [{\citenamefont {Yang}\ \emph {et~al.}(2014)\citenamefont {Yang},
  \citenamefont {Xu}, \citenamefont {Lu},\ and\ \citenamefont
  {Loh}}]{JAN_JACS136}%
  \BibitemOpen
  \bibfield  {author} {\bibinfo {author} {\bibfnamefont {B.}~\bibnamefont
  {Yang}}, \bibinfo {author} {\bibfnamefont {H.}~\bibnamefont {Xu}}, \bibinfo
  {author} {\bibfnamefont {J.}~\bibnamefont {Lu}}, \ and\ \bibinfo {author}
  {\bibfnamefont {K.~P.}\ \bibnamefont {Loh}},\ }\href {\doibase
  10.1021/ja5054847} {\bibfield  {journal} {\bibinfo  {journal} {J. Am. Chem.
  Soc.}\ }\textbf {\bibinfo {volume} {136}},\ \bibinfo {pages} {12041}
  (\bibinfo {year} {2014})}\BibitemShut {NoStop}%
\bibitem [{\citenamefont {Wu}\ \emph {et~al.}(2013)\citenamefont {Wu},
  \citenamefont {Hao}, \citenamefont {Jeong}, \citenamefont {Lee},
  \citenamefont {Chen}, \citenamefont {Jiang}, \citenamefont {Wu},
  \citenamefont {Piner}, \citenamefont {Kang},\ and\ \citenamefont
  {Ruoff}}]{WU_AM25}%
  \BibitemOpen
  \bibfield  {author} {\bibinfo {author} {\bibfnamefont {Y.}~\bibnamefont
  {Wu}}, \bibinfo {author} {\bibfnamefont {Y.}~\bibnamefont {Hao}}, \bibinfo
  {author} {\bibfnamefont {H.~Y.}\ \bibnamefont {Jeong}}, \bibinfo {author}
  {\bibfnamefont {Z.}~\bibnamefont {Lee}}, \bibinfo {author} {\bibfnamefont
  {S.}~\bibnamefont {Chen}}, \bibinfo {author} {\bibfnamefont {W.}~\bibnamefont
  {Jiang}}, \bibinfo {author} {\bibfnamefont {Q.}~\bibnamefont {Wu}}, \bibinfo
  {author} {\bibfnamefont {R.~D.}\ \bibnamefont {Piner}}, \bibinfo {author}
  {\bibfnamefont {J.}~\bibnamefont {Kang}}, \ and\ \bibinfo {author}
  {\bibfnamefont {R.~S.}\ \bibnamefont {Ruoff}},\ }\href {\doibase
  10.1002/adma.201302208} {\bibfield  {journal} {\bibinfo  {journal} {Adv.
  Mater.}\ }\textbf {\bibinfo {volume} {25}},\ \bibinfo {pages} {6744}
  (\bibinfo {year} {2013})}\BibitemShut {NoStop}%
\bibitem [{\citenamefont {Robertson}\ \emph {et~al.}(2012)\citenamefont
  {Robertson}, \citenamefont {Allen}, \citenamefont {Wu}, \citenamefont {He},
  \citenamefont {Olivier}, \citenamefont {Neethling}, \citenamefont
  {Kirkland},\ and\ \citenamefont {Warner}}]{ROB_NC3}%
  \BibitemOpen
  \bibfield  {author} {\bibinfo {author} {\bibfnamefont {A.~W.}\ \bibnamefont
  {Robertson}}, \bibinfo {author} {\bibfnamefont {C.~S.}\ \bibnamefont
  {Allen}}, \bibinfo {author} {\bibfnamefont {Y.~A.}\ \bibnamefont {Wu}},
  \bibinfo {author} {\bibfnamefont {K.}~\bibnamefont {He}}, \bibinfo {author}
  {\bibfnamefont {J.}~\bibnamefont {Olivier}}, \bibinfo {author} {\bibfnamefont
  {J.}~\bibnamefont {Neethling}}, \bibinfo {author} {\bibfnamefont {A.~I.}\
  \bibnamefont {Kirkland}}, \ and\ \bibinfo {author} {\bibfnamefont {J.~H.}\
  \bibnamefont {Warner}},\ }\href {\doibase 10.1038/ncomms2141} {\bibfield
  {journal} {\bibinfo  {journal} {Nat. Commun.}\ }\textbf {\bibinfo {volume}
  {3}},\ \bibinfo {pages} {1144} (\bibinfo {year} {2012})}\BibitemShut
  {NoStop}%
\bibitem [{\citenamefont {Chen}\ \emph {et~al.}(2014)\citenamefont {Chen},
  \citenamefont {Aut{\`{e}}s}, \citenamefont {Alem}, \citenamefont {Gargiulo},
  \citenamefont {Gautam}, \citenamefont {Linck}, \citenamefont {Kisielowski},
  \citenamefont {Yazyev}, \citenamefont {Louie},\ and\ \citenamefont
  {Zettl}}]{CHE_PRB89}%
  \BibitemOpen
  \bibfield  {author} {\bibinfo {author} {\bibfnamefont {J.-H.}\ \bibnamefont
  {Chen}}, \bibinfo {author} {\bibfnamefont {G.}~\bibnamefont {Aut{\`{e}}s}},
  \bibinfo {author} {\bibfnamefont {N.}~\bibnamefont {Alem}}, \bibinfo {author}
  {\bibfnamefont {F.}~\bibnamefont {Gargiulo}}, \bibinfo {author}
  {\bibfnamefont {A.}~\bibnamefont {Gautam}}, \bibinfo {author} {\bibfnamefont
  {M.}~\bibnamefont {Linck}}, \bibinfo {author} {\bibfnamefont
  {C.}~\bibnamefont {Kisielowski}}, \bibinfo {author} {\bibfnamefont {O.~V.}\
  \bibnamefont {Yazyev}}, \bibinfo {author} {\bibfnamefont {S.~G.}\
  \bibnamefont {Louie}}, \ and\ \bibinfo {author} {\bibfnamefont
  {A.}~\bibnamefont {Zettl}},\ }\href {\doibase 10.1103/physrevb.89.121407}
  {\bibfield  {journal} {\bibinfo  {journal} {Phys. Rev. B}\ }\textbf {\bibinfo
  {volume} {89}},\ \bibinfo {pages} {121407} (\bibinfo {year}
  {2014})}\BibitemShut {NoStop}%
\bibitem [{\citenamefont {de~Souza}\ \emph {et~al.}(2018)\citenamefont
  {de~Souza}, \citenamefont {Amorim}, \citenamefont {Prasongkit}, \citenamefont
  {Scopel}, \citenamefont {Scheicher},\ and\ \citenamefont
  {Rocha}}]{Souza2018}%
  \BibitemOpen
  \bibfield  {author} {\bibinfo {author} {\bibfnamefont {F.~A.}\ \bibnamefont
  {de~Souza}}, \bibinfo {author} {\bibfnamefont {R.~G.}\ \bibnamefont
  {Amorim}}, \bibinfo {author} {\bibfnamefont {J.}~\bibnamefont {Prasongkit}},
  \bibinfo {author} {\bibfnamefont {W.~L.}\ \bibnamefont {Scopel}}, \bibinfo
  {author} {\bibfnamefont {R.~H.}\ \bibnamefont {Scheicher}}, \ and\ \bibinfo
  {author} {\bibfnamefont {A.~R.}\ \bibnamefont {Rocha}},\ }\href {\doibase
  10.1016/j.carbon.2017.11.029} {\bibfield  {journal} {\bibinfo  {journal}
  {Carbon}\ }\textbf {\bibinfo {volume} {129}},\ \bibinfo {pages} {803}
  (\bibinfo {year} {2018})}\BibitemShut {NoStop}%
\bibitem [{\citenamefont {Veliev}\ \emph {et~al.}(2018)\citenamefont {Veliev},
  \citenamefont {Cresti}, \citenamefont {Kalita}, \citenamefont {Bourrier},
  \citenamefont {Belloir}, \citenamefont {Brian{\c{c}}on-Marjollet},
  \citenamefont {Albrieux}, \citenamefont {Roche}, \citenamefont {Bouchiat},\
  and\ \citenamefont {Delacour}}]{Veliev2018}%
  \BibitemOpen
  \bibfield  {author} {\bibinfo {author} {\bibfnamefont {F.}~\bibnamefont
  {Veliev}}, \bibinfo {author} {\bibfnamefont {A.}~\bibnamefont {Cresti}},
  \bibinfo {author} {\bibfnamefont {D.}~\bibnamefont {Kalita}}, \bibinfo
  {author} {\bibfnamefont {A.}~\bibnamefont {Bourrier}}, \bibinfo {author}
  {\bibfnamefont {T.}~\bibnamefont {Belloir}}, \bibinfo {author} {\bibfnamefont
  {A.}~\bibnamefont {Brian{\c{c}}on-Marjollet}}, \bibinfo {author}
  {\bibfnamefont {M.}~\bibnamefont {Albrieux}}, \bibinfo {author}
  {\bibfnamefont {S.}~\bibnamefont {Roche}}, \bibinfo {author} {\bibfnamefont
  {V.}~\bibnamefont {Bouchiat}}, \ and\ \bibinfo {author} {\bibfnamefont
  {C.}~\bibnamefont {Delacour}},\ }\href {\doibase 10.1088/2053-1583/aad78f}
  {\bibfield  {journal} {\bibinfo  {journal} {2D Mater.}\ }\textbf {\bibinfo
  {volume} {5}},\ \bibinfo {pages} {045020} (\bibinfo {year}
  {2018})}\BibitemShut {NoStop}%
\bibitem [{\citenamefont {Pop}\ \emph {et~al.}(2012)\citenamefont {Pop},
  \citenamefont {Varshney},\ and\ \citenamefont {Roy}}]{pop_thermal_2012}%
  \BibitemOpen
  \bibfield  {author} {\bibinfo {author} {\bibfnamefont {E.}~\bibnamefont
  {Pop}}, \bibinfo {author} {\bibfnamefont {V.}~\bibnamefont {Varshney}}, \
  and\ \bibinfo {author} {\bibfnamefont {A.~K.}\ \bibnamefont {Roy}},\ }\href
  {\doibase 10.1557/mrs.2012.203} {\bibfield  {journal} {\bibinfo  {journal}
  {MRS Bull.}\ }\textbf {\bibinfo {volume} {37}},\ \bibinfo {pages} {1273}
  (\bibinfo {year} {2012})}\BibitemShut {NoStop}%
\bibitem [{\citenamefont {Lindsay}\ \emph {et~al.}(2014)\citenamefont
  {Lindsay}, \citenamefont {Li}, \citenamefont {Carrete}, \citenamefont
  {Mingo}, \citenamefont {Broido},\ and\ \citenamefont
  {Reinecke}}]{lindsay_phonon_2014}%
  \BibitemOpen
  \bibfield  {author} {\bibinfo {author} {\bibfnamefont {L.}~\bibnamefont
  {Lindsay}}, \bibinfo {author} {\bibfnamefont {W.}~\bibnamefont {Li}},
  \bibinfo {author} {\bibfnamefont {J.}~\bibnamefont {Carrete}}, \bibinfo
  {author} {\bibfnamefont {N.}~\bibnamefont {Mingo}}, \bibinfo {author}
  {\bibfnamefont {D.~A.}\ \bibnamefont {Broido}}, \ and\ \bibinfo {author}
  {\bibfnamefont {T.~L.}\ \bibnamefont {Reinecke}},\ }\href {\doibase
  10.1103/physrevb.89.155426} {\bibfield  {journal} {\bibinfo  {journal} {Phys.
  Rev. B}\ }\textbf {\bibinfo {volume} {89}},\ \bibinfo {pages} {155426}
  (\bibinfo {year} {2014})}\BibitemShut {NoStop}%
\bibitem [{\citenamefont {Fugallo}\ \emph {et~al.}(2014)\citenamefont
  {Fugallo}, \citenamefont {Cepellotti}, \citenamefont {Paulatto},
  \citenamefont {Lazzeri}, \citenamefont {Marzari},\ and\ \citenamefont
  {Mauri}}]{Fugallo_NL14}%
  \BibitemOpen
  \bibfield  {author} {\bibinfo {author} {\bibfnamefont {G.}~\bibnamefont
  {Fugallo}}, \bibinfo {author} {\bibfnamefont {A.}~\bibnamefont {Cepellotti}},
  \bibinfo {author} {\bibfnamefont {L.}~\bibnamefont {Paulatto}}, \bibinfo
  {author} {\bibfnamefont {M.}~\bibnamefont {Lazzeri}}, \bibinfo {author}
  {\bibfnamefont {N.}~\bibnamefont {Marzari}}, \ and\ \bibinfo {author}
  {\bibfnamefont {F.}~\bibnamefont {Mauri}},\ }\href {\doibase
  10.1021/nl502059f} {\bibfield  {journal} {\bibinfo  {journal} {Nano Lett.}\
  }\textbf {\bibinfo {volume} {14}},\ \bibinfo {pages} {6109} (\bibinfo {year}
  {2014})}\BibitemShut {NoStop}%
\bibitem [{\citenamefont {Xu}\ \emph {et~al.}(2014)\citenamefont {Xu},
  \citenamefont {Pereira}, \citenamefont {Wang}, \citenamefont {Wu},
  \citenamefont {Zhang}, \citenamefont {Zhao}, \citenamefont {Bae},
  \citenamefont {Bui}, \citenamefont {Xie}, \citenamefont {Thong},
  \citenamefont {Hong}, \citenamefont {Loh}, \citenamefont {Donadio},
  \citenamefont {Li},\ and\ \citenamefont {Ã–zyilmaz}}]{Xu_NatComm14}%
  \BibitemOpen
  \bibfield  {author} {\bibinfo {author} {\bibfnamefont {X.}~\bibnamefont
  {Xu}}, \bibinfo {author} {\bibfnamefont {L.~F.~C.}\ \bibnamefont {Pereira}},
  \bibinfo {author} {\bibfnamefont {Y.}~\bibnamefont {Wang}}, \bibinfo {author}
  {\bibfnamefont {J.}~\bibnamefont {Wu}}, \bibinfo {author} {\bibfnamefont
  {K.}~\bibnamefont {Zhang}}, \bibinfo {author} {\bibfnamefont
  {X.}~\bibnamefont {Zhao}}, \bibinfo {author} {\bibfnamefont {S.}~\bibnamefont
  {Bae}}, \bibinfo {author} {\bibfnamefont {C.~T.}\ \bibnamefont {Bui}},
  \bibinfo {author} {\bibfnamefont {R.}~\bibnamefont {Xie}}, \bibinfo {author}
  {\bibfnamefont {J.~T.~L.}\ \bibnamefont {Thong}}, \bibinfo {author}
  {\bibfnamefont {B.~H.}\ \bibnamefont {Hong}}, \bibinfo {author}
  {\bibfnamefont {K.~P.}\ \bibnamefont {Loh}}, \bibinfo {author} {\bibfnamefont
  {D.}~\bibnamefont {Donadio}}, \bibinfo {author} {\bibfnamefont
  {B.}~\bibnamefont {Li}}, \ and\ \bibinfo {author} {\bibfnamefont
  {B.}~\bibnamefont {Ã–zyilmaz}},\ }\href {\doibase 10.1038/ncomms4689}
  {\bibfield  {journal} {\bibinfo  {journal} {Nat. Commun.}\ }\textbf {\bibinfo
  {volume} {5}},\ \bibinfo {pages} {3689} (\bibinfo {year} {2014})}\BibitemShut
  {NoStop}%
\bibitem [{\citenamefont {Lindsay}\ \emph {et~al.}(2010)\citenamefont
  {Lindsay}, \citenamefont {Broido},\ and\ \citenamefont
  {Mingo}}]{lindsay_flexural_2010}%
  \BibitemOpen
  \bibfield  {author} {\bibinfo {author} {\bibfnamefont {L.}~\bibnamefont
  {Lindsay}}, \bibinfo {author} {\bibfnamefont {D.~A.}\ \bibnamefont {Broido}},
  \ and\ \bibinfo {author} {\bibfnamefont {N.}~\bibnamefont {Mingo}},\ }\href
  {\doibase 10.1103/PhysRevB.82.115427} {\bibfield  {journal} {\bibinfo
  {journal} {Phys. Rev. B}\ }\textbf {\bibinfo {volume} {82}},\ \bibinfo
  {pages} {115427} (\bibinfo {year} {2010})}\BibitemShut {NoStop}%
\bibitem [{\citenamefont {Kolesnikov}\ and\ \citenamefont
  {Osipov}(2012)}]{KOL_EPL100}%
  \BibitemOpen
  \bibfield  {author} {\bibinfo {author} {\bibfnamefont {D.~V.}\ \bibnamefont
  {Kolesnikov}}\ and\ \bibinfo {author} {\bibfnamefont {V.~A.}\ \bibnamefont
  {Osipov}},\ }\href {\doibase 10.1209/0295-5075/100/26004} {\bibfield
  {journal} {\bibinfo  {journal} {{EPL} (Europhysics Letters)}\ }\textbf
  {\bibinfo {volume} {100}},\ \bibinfo {pages} {26004} (\bibinfo {year}
  {2012})}\BibitemShut {NoStop}%
\bibitem [{\citenamefont {Khosravian}\ \emph {et~al.}(2013)\citenamefont
  {Khosravian}, \citenamefont {Samani}, \citenamefont {Loh}, \citenamefont
  {Chen}, \citenamefont {Baillargeat},\ and\ \citenamefont {Tay}}]{KHO_CMS79}%
  \BibitemOpen
  \bibfield  {author} {\bibinfo {author} {\bibfnamefont {N.}~\bibnamefont
  {Khosravian}}, \bibinfo {author} {\bibfnamefont {M.}~\bibnamefont {Samani}},
  \bibinfo {author} {\bibfnamefont {G.}~\bibnamefont {Loh}}, \bibinfo {author}
  {\bibfnamefont {G.}~\bibnamefont {Chen}}, \bibinfo {author} {\bibfnamefont
  {D.}~\bibnamefont {Baillargeat}}, \ and\ \bibinfo {author} {\bibfnamefont
  {B.}~\bibnamefont {Tay}},\ }\href {\doibase 10.1016/j.commatsci.2013.06.002}
  {\bibfield  {journal} {\bibinfo  {journal} {Computational Materials Science}\
  }\textbf {\bibinfo {volume} {79}},\ \bibinfo {pages} {132} (\bibinfo {year}
  {2013})}\BibitemShut {NoStop}%
\bibitem [{\citenamefont {Krasavin}\ and\ \citenamefont
  {Osipov}(2015)}]{krasavin_effect_2015}%
  \BibitemOpen
  \bibfield  {author} {\bibinfo {author} {\bibfnamefont {S.~E.}\ \bibnamefont
  {Krasavin}}\ and\ \bibinfo {author} {\bibfnamefont {V.~A.}\ \bibnamefont
  {Osipov}},\ }\href {\doibase 10.1088/0953-8984/27/42/425302} {\bibfield
  {journal} {\bibinfo  {journal} {J. Phys.: Condens. Matter}\ }\textbf
  {\bibinfo {volume} {27}},\ \bibinfo {pages} {425302} (\bibinfo {year}
  {2015})}\BibitemShut {NoStop}%
\bibitem [{\citenamefont {Zhao}\ \emph {et~al.}(2015)\citenamefont {Zhao},
  \citenamefont {Wang}, \citenamefont {Wu}, \citenamefont {Wang}, \citenamefont
  {Bi}, \citenamefont {Liang}, \citenamefont {Yang}, \citenamefont {Chen},
  \citenamefont {Xu},\ and\ \citenamefont {Ni}}]{zhao_defect-engineered_2015}%
  \BibitemOpen
  \bibfield  {author} {\bibinfo {author} {\bibfnamefont {W.}~\bibnamefont
  {Zhao}}, \bibinfo {author} {\bibfnamefont {Y.}~\bibnamefont {Wang}}, \bibinfo
  {author} {\bibfnamefont {Z.}~\bibnamefont {Wu}}, \bibinfo {author}
  {\bibfnamefont {W.}~\bibnamefont {Wang}}, \bibinfo {author} {\bibfnamefont
  {K.}~\bibnamefont {Bi}}, \bibinfo {author} {\bibfnamefont {Z.}~\bibnamefont
  {Liang}}, \bibinfo {author} {\bibfnamefont {J.}~\bibnamefont {Yang}},
  \bibinfo {author} {\bibfnamefont {Y.}~\bibnamefont {Chen}}, \bibinfo {author}
  {\bibfnamefont {Z.}~\bibnamefont {Xu}}, \ and\ \bibinfo {author}
  {\bibfnamefont {Z.}~\bibnamefont {Ni}},\ }\href {\doibase 10.1038/srep11962}
  {\bibfield  {journal} {\bibinfo  {journal} {Sci. Rep.}\ }\textbf {\bibinfo
  {volume} {5}},\ \bibinfo {pages} {11962} (\bibinfo {year}
  {2015})}\BibitemShut {NoStop}%
\bibitem [{\citenamefont {Hahn}\ \emph {et~al.}(2016)\citenamefont {Hahn},
  \citenamefont {Melis},\ and\ \citenamefont {Colombo}}]{Hahn2016}%
  \BibitemOpen
  \bibfield  {author} {\bibinfo {author} {\bibfnamefont {K.~R.}\ \bibnamefont
  {Hahn}}, \bibinfo {author} {\bibfnamefont {C.}~\bibnamefont {Melis}}, \ and\
  \bibinfo {author} {\bibfnamefont {L.}~\bibnamefont {Colombo}},\ }\href
  {\doibase 10.1016/j.carbon.2015.09.070} {\bibfield  {journal} {\bibinfo
  {journal} {Carbon}\ }\textbf {\bibinfo {volume} {96}},\ \bibinfo {pages}
  {429} (\bibinfo {year} {2016})}\BibitemShut {NoStop}%
\bibitem [{\citenamefont {Sevim}\ and\ \citenamefont
  {Sevin{\c{c}}li}(2018)}]{Sevim2018}%
  \BibitemOpen
  \bibfield  {author} {\bibinfo {author} {\bibfnamefont {K.}~\bibnamefont
  {Sevim}}\ and\ \bibinfo {author} {\bibfnamefont {H.}~\bibnamefont
  {Sevin{\c{c}}li}},\ }\href {\doibase 10.1016/j.carbon.2018.08.050} {\bibfield
   {journal} {\bibinfo  {journal} {Carbon}\ }\textbf {\bibinfo {volume}
  {140}},\ \bibinfo {pages} {603} (\bibinfo {year} {2018})}\BibitemShut
  {NoStop}%
\bibitem [{\citenamefont {Nobakht}\ \emph {et~al.}(2018)\citenamefont
  {Nobakht}, \citenamefont {Gandomi}, \citenamefont {Wang}, \citenamefont
  {Bowman}, \citenamefont {Marable}, \citenamefont {Garrison}, \citenamefont
  {Kim},\ and\ \citenamefont {Shin}}]{Nobakht2018}%
  \BibitemOpen
  \bibfield  {author} {\bibinfo {author} {\bibfnamefont {A.~Y.}\ \bibnamefont
  {Nobakht}}, \bibinfo {author} {\bibfnamefont {Y.~A.}\ \bibnamefont
  {Gandomi}}, \bibinfo {author} {\bibfnamefont {J.}~\bibnamefont {Wang}},
  \bibinfo {author} {\bibfnamefont {M.~H.}\ \bibnamefont {Bowman}}, \bibinfo
  {author} {\bibfnamefont {D.~C.}\ \bibnamefont {Marable}}, \bibinfo {author}
  {\bibfnamefont {B.~E.}\ \bibnamefont {Garrison}}, \bibinfo {author}
  {\bibfnamefont {D.}~\bibnamefont {Kim}}, \ and\ \bibinfo {author}
  {\bibfnamefont {S.}~\bibnamefont {Shin}},\ }\href {\doibase
  10.1016/j.carbon.2018.02.087} {\bibfield  {journal} {\bibinfo  {journal}
  {Carbon}\ }\textbf {\bibinfo {volume} {132}},\ \bibinfo {pages} {565}
  (\bibinfo {year} {2018})}\BibitemShut {NoStop}%
\bibitem [{\citenamefont {Tan}\ \emph {et~al.}(2013)\citenamefont {Tan},
  \citenamefont {Tang}, \citenamefont {Xie}, \citenamefont {Pan},\ and\
  \citenamefont {Chen}}]{TAN_CAR65}%
  \BibitemOpen
  \bibfield  {author} {\bibinfo {author} {\bibfnamefont {S.-H.}\ \bibnamefont
  {Tan}}, \bibinfo {author} {\bibfnamefont {L.-M.}\ \bibnamefont {Tang}},
  \bibinfo {author} {\bibfnamefont {Z.-X.}\ \bibnamefont {Xie}}, \bibinfo
  {author} {\bibfnamefont {C.-N.}\ \bibnamefont {Pan}}, \ and\ \bibinfo
  {author} {\bibfnamefont {K.-Q.}\ \bibnamefont {Chen}},\ }\href {\doibase
  10.1016/j.carbon.2013.08.012} {\bibfield  {journal} {\bibinfo  {journal}
  {Carbon}\ }\textbf {\bibinfo {volume} {65}},\ \bibinfo {pages} {181}
  (\bibinfo {year} {2013})}\BibitemShut {NoStop}%
\bibitem [{\citenamefont {Yang}\ \emph {et~al.}(2013)\citenamefont {Yang},
  \citenamefont {Li}, \citenamefont {Zhao}, \citenamefont {Yang},\ and\
  \citenamefont {Wang}}]{YAN_PLA377}%
  \BibitemOpen
  \bibfield  {author} {\bibinfo {author} {\bibfnamefont {P.}~\bibnamefont
  {Yang}}, \bibinfo {author} {\bibfnamefont {X.}~\bibnamefont {Li}}, \bibinfo
  {author} {\bibfnamefont {Y.}~\bibnamefont {Zhao}}, \bibinfo {author}
  {\bibfnamefont {H.}~\bibnamefont {Yang}}, \ and\ \bibinfo {author}
  {\bibfnamefont {S.}~\bibnamefont {Wang}},\ }\href {\doibase
  10.1016/j.physleta.2013.05.058} {\bibfield  {journal} {\bibinfo  {journal}
  {Phys. Lett. A}\ }\textbf {\bibinfo {volume} {377}},\ \bibinfo {pages} {2141}
  (\bibinfo {year} {2013})}\BibitemShut {NoStop}%
\bibitem [{\citenamefont {Zhu}\ \emph {et~al.}(2016)\citenamefont {Zhu},
  \citenamefont {Yang}, \citenamefont {Huang}, \citenamefont {He},
  \citenamefont {Yang},\ and\ \citenamefont {Wang}}]{zhu_mechanisms_2016}%
  \BibitemOpen
  \bibfield  {author} {\bibinfo {author} {\bibfnamefont {Z.}~\bibnamefont
  {Zhu}}, \bibinfo {author} {\bibfnamefont {X.}~\bibnamefont {Yang}}, \bibinfo
  {author} {\bibfnamefont {M.}~\bibnamefont {Huang}}, \bibinfo {author}
  {\bibfnamefont {Q.}~\bibnamefont {He}}, \bibinfo {author} {\bibfnamefont
  {G.}~\bibnamefont {Yang}}, \ and\ \bibinfo {author} {\bibfnamefont
  {Z.}~\bibnamefont {Wang}},\ }\href {\doibase 10.1088/0957-4484/27/5/055401}
  {\bibfield  {journal} {\bibinfo  {journal} {Nanotechnology}\ }\textbf
  {\bibinfo {volume} {27}},\ \bibinfo {pages} {055401} (\bibinfo {year}
  {2016})}\BibitemShut {NoStop}%
\bibitem [{\citenamefont {Yan}\ \emph {et~al.}(2013)\citenamefont {Yan},
  \citenamefont {Liu}, \citenamefont {Bai}, \citenamefont {Wang}, \citenamefont
  {Liu}, \citenamefont {Yan}, \citenamefont {Meng}, \citenamefont {Zhang},
  \citenamefont {Liu}, \citenamefont {fen Dou}, \citenamefont {Nie},
  \citenamefont {Yao},\ and\ \citenamefont {He}}]{YAN_APL103}%
  \BibitemOpen
  \bibfield  {author} {\bibinfo {author} {\bibfnamefont {H.}~\bibnamefont
  {Yan}}, \bibinfo {author} {\bibfnamefont {C.-C.}\ \bibnamefont {Liu}},
  \bibinfo {author} {\bibfnamefont {K.-K.}\ \bibnamefont {Bai}}, \bibinfo
  {author} {\bibfnamefont {X.}~\bibnamefont {Wang}}, \bibinfo {author}
  {\bibfnamefont {M.}~\bibnamefont {Liu}}, \bibinfo {author} {\bibfnamefont
  {W.}~\bibnamefont {Yan}}, \bibinfo {author} {\bibfnamefont {L.}~\bibnamefont
  {Meng}}, \bibinfo {author} {\bibfnamefont {Y.}~\bibnamefont {Zhang}},
  \bibinfo {author} {\bibfnamefont {Z.}~\bibnamefont {Liu}}, \bibinfo {author}
  {\bibfnamefont {R.}~\bibnamefont {fen Dou}}, \bibinfo {author} {\bibfnamefont
  {J.-C.}\ \bibnamefont {Nie}}, \bibinfo {author} {\bibfnamefont
  {Y.}~\bibnamefont {Yao}}, \ and\ \bibinfo {author} {\bibfnamefont
  {L.}~\bibnamefont {He}},\ }\href {\doibase 10.1063/1.4824206} {\bibfield
  {journal} {\bibinfo  {journal} {Appl. Phys. Lett.}\ }\textbf {\bibinfo
  {volume} {103}},\ \bibinfo {pages} {143120} (\bibinfo {year}
  {2013})}\BibitemShut {NoStop}%
\bibitem [{\citenamefont {Kang}\ \emph {et~al.}(2019)\citenamefont {Kang},
  \citenamefont {Zhang},\ and\ \citenamefont {Li}}]{Kang2019}%
  \BibitemOpen
  \bibfield  {author} {\bibinfo {author} {\bibfnamefont {D.}~\bibnamefont
  {Kang}}, \bibinfo {author} {\bibfnamefont {C.}~\bibnamefont {Zhang}}, \ and\
  \bibinfo {author} {\bibfnamefont {H.}~\bibnamefont {Li}},\ }\href {\doibase
  10.1007/s10948-019-05180-y} {\bibfield  {journal} {\bibinfo  {journal} {J.
  Supercond. Novel Magn.}\ } (\bibinfo {year} {2019}),\
  10.1007/s10948-019-05180-y}\BibitemShut {NoStop}%
\bibitem [{\citenamefont {Tyurnina}\ \emph {et~al.}(2016)\citenamefont
  {Tyurnina}, \citenamefont {Okuno}, \citenamefont {Pochet},\ and\
  \citenamefont {Dijon}}]{TYU_CA102}%
  \BibitemOpen
  \bibfield  {author} {\bibinfo {author} {\bibfnamefont {A.~V.}\ \bibnamefont
  {Tyurnina}}, \bibinfo {author} {\bibfnamefont {H.}~\bibnamefont {Okuno}},
  \bibinfo {author} {\bibfnamefont {P.}~\bibnamefont {Pochet}}, \ and\ \bibinfo
  {author} {\bibfnamefont {J.}~\bibnamefont {Dijon}},\ }\href {\doibase
  10.1016/j.carbon.2016.02.097} {\bibfield  {journal} {\bibinfo  {journal}
  {Carbon}\ }\textbf {\bibinfo {volume} {102}},\ \bibinfo {pages} {499}
  (\bibinfo {year} {2016})}\BibitemShut {NoStop}%
\bibitem [{\citenamefont {Zhao}\ \emph {et~al.}(2013)\citenamefont {Zhao},
  \citenamefont {Levendorf}, \citenamefont {Goncher}, \citenamefont {Schiros},
  \citenamefont {P{\'{a}}lov{\'{a}}}, \citenamefont {Zabet-Khosousi},
  \citenamefont {Rim}, \citenamefont {Guti{\'{e}}rrez}, \citenamefont
  {Nordlund}, \citenamefont {Jaye}, \citenamefont {Hybertsen}, \citenamefont
  {Reichman}, \citenamefont {Flynn}, \citenamefont {Park},\ and\ \citenamefont
  {Pasupathy}}]{Zhao_NL2013}%
  \BibitemOpen
  \bibfield  {author} {\bibinfo {author} {\bibfnamefont {L.}~\bibnamefont
  {Zhao}}, \bibinfo {author} {\bibfnamefont {M.}~\bibnamefont {Levendorf}},
  \bibinfo {author} {\bibfnamefont {S.}~\bibnamefont {Goncher}}, \bibinfo
  {author} {\bibfnamefont {T.}~\bibnamefont {Schiros}}, \bibinfo {author}
  {\bibfnamefont {L.}~\bibnamefont {P{\'{a}}lov{\'{a}}}}, \bibinfo {author}
  {\bibfnamefont {A.}~\bibnamefont {Zabet-Khosousi}}, \bibinfo {author}
  {\bibfnamefont {K.~T.}\ \bibnamefont {Rim}}, \bibinfo {author} {\bibfnamefont
  {C.}~\bibnamefont {Guti{\'{e}}rrez}}, \bibinfo {author} {\bibfnamefont
  {D.}~\bibnamefont {Nordlund}}, \bibinfo {author} {\bibfnamefont
  {C.}~\bibnamefont {Jaye}}, \bibinfo {author} {\bibfnamefont {M.}~\bibnamefont
  {Hybertsen}}, \bibinfo {author} {\bibfnamefont {D.}~\bibnamefont {Reichman}},
  \bibinfo {author} {\bibfnamefont {G.~W.}\ \bibnamefont {Flynn}}, \bibinfo
  {author} {\bibfnamefont {J.}~\bibnamefont {Park}}, \ and\ \bibinfo {author}
  {\bibfnamefont {A.~N.}\ \bibnamefont {Pasupathy}},\ }\href {\doibase
  10.1021/nl401781d} {\bibfield  {journal} {\bibinfo  {journal} {Nano Lett.}\
  }\textbf {\bibinfo {volume} {13}},\ \bibinfo {pages} {4659} (\bibinfo {year}
  {2013})}\BibitemShut {NoStop}%
\bibitem [{\citenamefont {Cui}\ \emph {et~al.}(2017)\citenamefont {Cui},
  \citenamefont {Zhang}, \citenamefont {Chen}, \citenamefont {Yang},\ and\
  \citenamefont {Cai}}]{CUI_JPCC2017}%
  \BibitemOpen
  \bibfield  {author} {\bibinfo {author} {\bibfnamefont {Y.}~\bibnamefont
  {Cui}}, \bibinfo {author} {\bibfnamefont {H.}~\bibnamefont {Zhang}}, \bibinfo
  {author} {\bibfnamefont {W.}~\bibnamefont {Chen}}, \bibinfo {author}
  {\bibfnamefont {Z.}~\bibnamefont {Yang}}, \ and\ \bibinfo {author}
  {\bibfnamefont {Q.}~\bibnamefont {Cai}},\ }\href {\doibase
  10.1021/acs.jpcc.7b04693} {\bibfield  {journal} {\bibinfo  {journal} {The
  Journal of Physical Chemistry C}\ }\textbf {\bibinfo {volume} {121}},\
  \bibinfo {pages} {15282} (\bibinfo {year} {2017})}\BibitemShut {NoStop}%
\bibitem [{\citenamefont {Wang}\ \emph {et~al.}(2016)\citenamefont {Wang},
  \citenamefont {Dong}, \citenamefont {Cui}, \citenamefont {Eres},
  \citenamefont {Timpe}, \citenamefont {Fu}, \citenamefont {Ding},
  \citenamefont {Schloegl},\ and\ \citenamefont {Willinger}}]{Wang2016}%
  \BibitemOpen
  \bibfield  {author} {\bibinfo {author} {\bibfnamefont {Z.-J.}\ \bibnamefont
  {Wang}}, \bibinfo {author} {\bibfnamefont {J.}~\bibnamefont {Dong}}, \bibinfo
  {author} {\bibfnamefont {Y.}~\bibnamefont {Cui}}, \bibinfo {author}
  {\bibfnamefont {G.}~\bibnamefont {Eres}}, \bibinfo {author} {\bibfnamefont
  {O.}~\bibnamefont {Timpe}}, \bibinfo {author} {\bibfnamefont
  {Q.}~\bibnamefont {Fu}}, \bibinfo {author} {\bibfnamefont {F.}~\bibnamefont
  {Ding}}, \bibinfo {author} {\bibfnamefont {R.}~\bibnamefont {Schloegl}}, \
  and\ \bibinfo {author} {\bibfnamefont {M.-G.}\ \bibnamefont {Willinger}},\
  }\href {\doibase 10.1038/ncomms13256} {\bibfield  {journal} {\bibinfo
  {journal} {Nature Communications}\ }\textbf {\bibinfo {volume} {7}},\
  \bibinfo {pages} {13256} (\bibinfo {year} {2016})}\BibitemShut {NoStop}%
\bibitem [{\citenamefont {Rios}\ \emph {et~al.}(2005)\citenamefont {Rios},
  \citenamefont {Jr}, \citenamefont {Sandim}, \citenamefont {Plaut},\ and\
  \citenamefont {Padilha}}]{RIOS2006}%
  \BibitemOpen
  \bibfield  {author} {\bibinfo {author} {\bibfnamefont {P.~R.}\ \bibnamefont
  {Rios}}, \bibinfo {author} {\bibfnamefont {F.~S.}\ \bibnamefont {Jr}},
  \bibinfo {author} {\bibfnamefont {H.~R.~Z.}\ \bibnamefont {Sandim}}, \bibinfo
  {author} {\bibfnamefont {R.~L.}\ \bibnamefont {Plaut}}, \ and\ \bibinfo
  {author} {\bibfnamefont {A.~F.}\ \bibnamefont {Padilha}},\ }\href {\doibase
  10.1590/s1516-14392005000300002} {\bibfield  {journal} {\bibinfo  {journal}
  {Mater. Res.}\ }\textbf {\bibinfo {volume} {8}},\ \bibinfo {pages} {225}
  (\bibinfo {year} {2005})}\BibitemShut {NoStop}%
\bibitem [{\citenamefont {Chauve}\ \emph {et~al.}(2017)\citenamefont {Chauve},
  \citenamefont {Montagnat}, \citenamefont {Barou}, \citenamefont {Hidas},
  \citenamefont {Tommasi},\ and\ \citenamefont {Mainprice}}]{CHAUVE2016}%
  \BibitemOpen
  \bibfield  {author} {\bibinfo {author} {\bibfnamefont {T.}~\bibnamefont
  {Chauve}}, \bibinfo {author} {\bibfnamefont {M.}~\bibnamefont {Montagnat}},
  \bibinfo {author} {\bibfnamefont {F.}~\bibnamefont {Barou}}, \bibinfo
  {author} {\bibfnamefont {K.}~\bibnamefont {Hidas}}, \bibinfo {author}
  {\bibfnamefont {A.}~\bibnamefont {Tommasi}}, \ and\ \bibinfo {author}
  {\bibfnamefont {D.}~\bibnamefont {Mainprice}},\ }\href {\doibase
  10.1098/rsta.2015.0345} {\bibfield  {journal} {\bibinfo  {journal}
  {Philosophical Transactions of the Royal Society A: Mathematical, Physical
  and Engineering Sciences}\ }\textbf {\bibinfo {volume} {375}},\ \bibinfo
  {pages} {20150345} (\bibinfo {year} {2017})}\BibitemShut {NoStop}%
\bibitem [{\citenamefont {Halfpenny}\ \emph {et~al.}(2006)\citenamefont
  {Halfpenny}, \citenamefont {Prior},\ and\ \citenamefont
  {Wheeler}}]{HALFPENNY2006}%
  \BibitemOpen
  \bibfield  {author} {\bibinfo {author} {\bibfnamefont {A.}~\bibnamefont
  {Halfpenny}}, \bibinfo {author} {\bibfnamefont {D.~J.}\ \bibnamefont
  {Prior}}, \ and\ \bibinfo {author} {\bibfnamefont {J.}~\bibnamefont
  {Wheeler}},\ }\href {\doibase https://doi.org/10.1016/j.tecto.2006.05.016}
  {\bibfield  {journal} {\bibinfo  {journal} {Tectonophysics}\ }\textbf
  {\bibinfo {volume} {427}},\ \bibinfo {pages} {3} (\bibinfo {year}
  {2006})}\BibitemShut {NoStop}%
\bibitem [{\citenamefont {Annevelink}\ \emph {et~al.}(2019)\citenamefont
  {Annevelink}, \citenamefont {Ertekin},\ and\ \citenamefont
  {Johnson}}]{emil2019}%
  \BibitemOpen
  \bibfield  {author} {\bibinfo {author} {\bibfnamefont {E.}~\bibnamefont
  {Annevelink}}, \bibinfo {author} {\bibfnamefont {E.}~\bibnamefont {Ertekin}},
  \ and\ \bibinfo {author} {\bibfnamefont {H.~T.}\ \bibnamefont {Johnson}},\
  }\href {\doibase 10.1016/j.actamat.2018.12.030} {\bibfield  {journal}
  {\bibinfo  {journal} {Acta Mater.}\ }\textbf {\bibinfo {volume} {166}},\
  \bibinfo {pages} {67} (\bibinfo {year} {2019})}\BibitemShut {NoStop}%
\bibitem [{\citenamefont {Stone}\ and\ \citenamefont
  {Wales}(1986)}]{stone_theoretical_1986}%
  \BibitemOpen
  \bibfield  {author} {\bibinfo {author} {\bibfnamefont {A.~J.}\ \bibnamefont
  {Stone}}\ and\ \bibinfo {author} {\bibfnamefont {D.~J.}\ \bibnamefont
  {Wales}},\ }\href {\doibase 10.1016/0009-2614(86)80661-3} {\bibfield
  {journal} {\bibinfo  {journal} {Chem. Phys. Lett.}\ }\textbf {\bibinfo
  {volume} {128}},\ \bibinfo {pages} {501} (\bibinfo {year}
  {1986})}\BibitemShut {NoStop}%
\bibitem [{\citenamefont {Avriller}\ \emph {et~al.}(2007)\citenamefont
  {Avriller}, \citenamefont {Roche}, \citenamefont {Triozon}, \citenamefont
  {Blase},\ and\ \citenamefont {Latil}}]{AVR_MPLB21}%
  \BibitemOpen
  \bibfield  {author} {\bibinfo {author} {\bibfnamefont {R.}~\bibnamefont
  {Avriller}}, \bibinfo {author} {\bibfnamefont {S.}~\bibnamefont {Roche}},
  \bibinfo {author} {\bibfnamefont {F.}~\bibnamefont {Triozon}}, \bibinfo
  {author} {\bibfnamefont {X.}~\bibnamefont {Blase}}, \ and\ \bibinfo {author}
  {\bibfnamefont {S.}~\bibnamefont {Latil}},\ }\href {\doibase
  10.1142/s0217984907014322} {\bibfield  {journal} {\bibinfo  {journal} {Modern
  Physics Letters B}\ }\textbf {\bibinfo {volume} {21}},\ \bibinfo {pages}
  {1955} (\bibinfo {year} {2007})}\BibitemShut {NoStop}%
\bibitem [{\citenamefont {Ma}\ \emph {et~al.}(2014)\citenamefont {Ma},
  \citenamefont {Sun}, \citenamefont {Zhao}, \citenamefont {Li}, \citenamefont
  {Li}, \citenamefont {Zhao}, \citenamefont {Wang}, \citenamefont {Luo},
  \citenamefont {Yang}, \citenamefont {Wang},\ and\ \citenamefont
  {Hou}}]{Ma2014}%
  \BibitemOpen
  \bibfield  {author} {\bibinfo {author} {\bibfnamefont {C.}~\bibnamefont
  {Ma}}, \bibinfo {author} {\bibfnamefont {H.}~\bibnamefont {Sun}}, \bibinfo
  {author} {\bibfnamefont {Y.}~\bibnamefont {Zhao}}, \bibinfo {author}
  {\bibfnamefont {B.}~\bibnamefont {Li}}, \bibinfo {author} {\bibfnamefont
  {Q.}~\bibnamefont {Li}}, \bibinfo {author} {\bibfnamefont {A.}~\bibnamefont
  {Zhao}}, \bibinfo {author} {\bibfnamefont {X.}~\bibnamefont {Wang}}, \bibinfo
  {author} {\bibfnamefont {Y.}~\bibnamefont {Luo}}, \bibinfo {author}
  {\bibfnamefont {J.}~\bibnamefont {Yang}}, \bibinfo {author} {\bibfnamefont
  {B.}~\bibnamefont {Wang}}, \ and\ \bibinfo {author} {\bibfnamefont
  {J.}~\bibnamefont {Hou}},\ }\href {\doibase 10.1103/physrevlett.112.226802}
  {\bibfield  {journal} {\bibinfo  {journal} {Phys. Rev. Lett.}\ }\textbf
  {\bibinfo {volume} {112}},\ \bibinfo {pages} {226802} (\bibinfo {year}
  {2014})}\BibitemShut {NoStop}%
\bibitem [{\citenamefont {Mucciolo}\ \emph {et~al.}(2009)\citenamefont
  {Mucciolo}, \citenamefont {Neto},\ and\ \citenamefont
  {Lewenkopf}}]{MUC_PRB79}%
  \BibitemOpen
  \bibfield  {author} {\bibinfo {author} {\bibfnamefont {E.~R.}\ \bibnamefont
  {Mucciolo}}, \bibinfo {author} {\bibfnamefont {A.~H.~C.}\ \bibnamefont
  {Neto}}, \ and\ \bibinfo {author} {\bibfnamefont {C.~H.}\ \bibnamefont
  {Lewenkopf}},\ }\href {\doibase 10.1103/physrevb.79.075407} {\bibfield
  {journal} {\bibinfo  {journal} {Phys. Rev. B}\ }\textbf {\bibinfo {volume}
  {79}},\ \bibinfo {pages} {075407} (\bibinfo {year} {2009})}\BibitemShut
  {NoStop}%
\bibitem [{\citenamefont {Dubois}\ \emph {et~al.}(2010)\citenamefont {Dubois},
  \citenamefont {Lopez-Bezanilla}, \citenamefont {Cresti}, \citenamefont
  {Triozon}, \citenamefont {Biel}, \citenamefont {Charlier},\ and\
  \citenamefont {Roche}}]{DUB_ACSN4}%
  \BibitemOpen
  \bibfield  {author} {\bibinfo {author} {\bibfnamefont {S.~M.-M.}\
  \bibnamefont {Dubois}}, \bibinfo {author} {\bibfnamefont {A.}~\bibnamefont
  {Lopez-Bezanilla}}, \bibinfo {author} {\bibfnamefont {A.}~\bibnamefont
  {Cresti}}, \bibinfo {author} {\bibfnamefont {F.}~\bibnamefont {Triozon}},
  \bibinfo {author} {\bibfnamefont {B.}~\bibnamefont {Biel}}, \bibinfo {author}
  {\bibfnamefont {J.-C.}\ \bibnamefont {Charlier}}, \ and\ \bibinfo {author}
  {\bibfnamefont {S.}~\bibnamefont {Roche}},\ }\href {\doibase
  10.1021/nn100028q} {\bibfield  {journal} {\bibinfo  {journal} {{ACS} Nano}\
  }\textbf {\bibinfo {volume} {4}},\ \bibinfo {pages} {1971} (\bibinfo {year}
  {2010})}\BibitemShut {NoStop}%
\bibitem [{\citenamefont {Saloriutta}\ \emph {et~al.}(2011)\citenamefont
  {Saloriutta}, \citenamefont {Hancock}, \citenamefont {K\"arkk\"ainen},
  \citenamefont {K\"arkk\"ainen}, \citenamefont {Puska},\ and\ \citenamefont
  {Jauho}}]{SAL_PRB83}%
  \BibitemOpen
  \bibfield  {author} {\bibinfo {author} {\bibfnamefont {K.}~\bibnamefont
  {Saloriutta}}, \bibinfo {author} {\bibfnamefont {Y.}~\bibnamefont {Hancock}},
  \bibinfo {author} {\bibfnamefont {A.}~\bibnamefont {K\"arkk\"ainen}},
  \bibinfo {author} {\bibfnamefont {L.}~\bibnamefont {K\"arkk\"ainen}},
  \bibinfo {author} {\bibfnamefont {M.~J.}\ \bibnamefont {Puska}}, \ and\
  \bibinfo {author} {\bibfnamefont {A.-P.}\ \bibnamefont {Jauho}},\ }\href
  {\doibase 10.1103/PhysRevB.83.205125} {\bibfield  {journal} {\bibinfo
  {journal} {Phys. Rev. B}\ }\textbf {\bibinfo {volume} {83}},\ \bibinfo
  {pages} {205125} (\bibinfo {year} {2011})}\BibitemShut {NoStop}%
\bibitem [{\citenamefont {Ihnatsenka}\ and\ \citenamefont
  {Kirczenow}(2013)}]{IHN_PRB88}%
  \BibitemOpen
  \bibfield  {author} {\bibinfo {author} {\bibfnamefont {S.}~\bibnamefont
  {Ihnatsenka}}\ and\ \bibinfo {author} {\bibfnamefont {G.}~\bibnamefont
  {Kirczenow}},\ }\href {\doibase 10.1103/physrevb.88.125430} {\bibfield
  {journal} {\bibinfo  {journal} {Phys. Rev. B}\ }\textbf {\bibinfo {volume}
  {88}},\ \bibinfo {pages} {125430} (\bibinfo {year} {2013})}\BibitemShut
  {NoStop}%
\bibitem [{\citenamefont {Li}\ \emph {et~al.}(2014)\citenamefont {Li},
  \citenamefont {Carrete}, \citenamefont {Katcho},\ and\ \citenamefont
  {Mingo}}]{cpc_2014}%
  \BibitemOpen
  \bibfield  {author} {\bibinfo {author} {\bibfnamefont {W.}~\bibnamefont
  {Li}}, \bibinfo {author} {\bibfnamefont {J.}~\bibnamefont {Carrete}},
  \bibinfo {author} {\bibfnamefont {N.~A.}\ \bibnamefont {Katcho}}, \ and\
  \bibinfo {author} {\bibfnamefont {N.}~\bibnamefont {Mingo}},\ }\href
  {\doibase 10.1016/j.cpc.2014.02.015} {\bibfield  {journal} {\bibinfo
  {journal} {Comp. Phys. Comm.}\ }\textbf {\bibinfo {volume} {185}},\ \bibinfo
  {pages} {1747} (\bibinfo {year} {2014})}\BibitemShut {NoStop}%
\bibitem [{\citenamefont {Katre}\ \emph {et~al.}(2016)\citenamefont {Katre},
  \citenamefont {Carrete},\ and\ \citenamefont
  {Mingo}}]{katre_unraveling_2016}%
  \BibitemOpen
  \bibfield  {author} {\bibinfo {author} {\bibfnamefont {A.}~\bibnamefont
  {Katre}}, \bibinfo {author} {\bibfnamefont {J.}~\bibnamefont {Carrete}}, \
  and\ \bibinfo {author} {\bibfnamefont {N.}~\bibnamefont {Mingo}},\ }\href
  {\doibase 10.1039/c6ta05868j} {\bibfield  {journal} {\bibinfo  {journal} {J.
  Mater. Chem. A}\ }\textbf {\bibinfo {volume} {4}},\ \bibinfo {pages} {15940}
  (\bibinfo {year} {2016})}\BibitemShut {NoStop}%
\bibitem [{\citenamefont {Wang}\ \emph {et~al.}(2017)\citenamefont {Wang},
  \citenamefont {Carrete}, \citenamefont {van Roekeghem}, \citenamefont
  {Mingo},\ and\ \citenamefont {Madsen}}]{Wang_PRB17}%
  \BibitemOpen
  \bibfield  {author} {\bibinfo {author} {\bibfnamefont {T.}~\bibnamefont
  {Wang}}, \bibinfo {author} {\bibfnamefont {J.}~\bibnamefont {Carrete}},
  \bibinfo {author} {\bibfnamefont {A.}~\bibnamefont {van Roekeghem}}, \bibinfo
  {author} {\bibfnamefont {N.}~\bibnamefont {Mingo}}, \ and\ \bibinfo {author}
  {\bibfnamefont {G.~K.~H.}\ \bibnamefont {Madsen}},\ }\href {\doibase
  10.1103/PhysRevB.95.245304} {\bibfield  {journal} {\bibinfo  {journal} {Phys.
  Rev. B}\ }\textbf {\bibinfo {volume} {95}},\ \bibinfo {pages} {245304}
  (\bibinfo {year} {2017})}\BibitemShut {NoStop}%
\bibitem [{\citenamefont {Carrete}\ \emph {et~al.}(2017)\citenamefont
  {Carrete}, \citenamefont {Vermeersch}, \citenamefont {Katre}, \citenamefont
  {van Roekeghem}, \citenamefont {Wang}, \citenamefont {Madsen},\ and\
  \citenamefont {Mingo}}]{almaBTE}%
  \BibitemOpen
  \bibfield  {author} {\bibinfo {author} {\bibfnamefont {J.}~\bibnamefont
  {Carrete}}, \bibinfo {author} {\bibfnamefont {B.}~\bibnamefont {Vermeersch}},
  \bibinfo {author} {\bibfnamefont {A.}~\bibnamefont {Katre}}, \bibinfo
  {author} {\bibfnamefont {A.}~\bibnamefont {van Roekeghem}}, \bibinfo {author}
  {\bibfnamefont {T.}~\bibnamefont {Wang}}, \bibinfo {author} {\bibfnamefont
  {G.~K.~H.}\ \bibnamefont {Madsen}}, \ and\ \bibinfo {author} {\bibfnamefont
  {N.}~\bibnamefont {Mingo}},\ }\href {\doibase 10.1016/j.cpc.2017.06.023}
  {\bibfield  {journal} {\bibinfo  {journal} {Comput. Phys. Commun.}\ }\textbf
  {\bibinfo {volume} {220}},\ \bibinfo {pages} {351} (\bibinfo {year}
  {2017})}\BibitemShut {NoStop}%
\bibitem [{\citenamefont {Lindsay}\ and\ \citenamefont
  {Broido}(2010)}]{lindsay_optimized_2010}%
  \BibitemOpen
  \bibfield  {author} {\bibinfo {author} {\bibfnamefont {L.}~\bibnamefont
  {Lindsay}}\ and\ \bibinfo {author} {\bibfnamefont {D.~A.}\ \bibnamefont
  {Broido}},\ }\href {\doibase 10.1103/PhysRevB.81.205441} {\bibfield
  {journal} {\bibinfo  {journal} {Phys. Rev. B}\ }\textbf {\bibinfo {volume}
  {81}},\ \bibinfo {pages} {205441} (\bibinfo {year} {2010})}\BibitemShut
  {NoStop}%
\bibitem [{\citenamefont {Wang}\ \emph {et~al.}(2019)\citenamefont {Wang},
  \citenamefont {Carrete}, \citenamefont {Mingo},\ and\ \citenamefont
  {Madsen}}]{dislocations2019}%
  \BibitemOpen
  \bibfield  {author} {\bibinfo {author} {\bibfnamefont {T.}~\bibnamefont
  {Wang}}, \bibinfo {author} {\bibfnamefont {J.}~\bibnamefont {Carrete}},
  \bibinfo {author} {\bibfnamefont {N.}~\bibnamefont {Mingo}}, \ and\ \bibinfo
  {author} {\bibfnamefont {G.~K.~H.}\ \bibnamefont {Madsen}},\ }\href {\doibase
  10.1021/acsami.8b17525} {\bibfield  {journal} {\bibinfo  {journal} {ACS Appl.
  Mater. Interfaces}\ }\textbf {\bibinfo {volume} {11}},\ \bibinfo {pages}
  {8175} (\bibinfo {year} {2019})}\BibitemShut {NoStop}%
\bibitem [{\citenamefont {Carrete}\ \emph {et~al.}(2016)\citenamefont
  {Carrete}, \citenamefont {Li}, \citenamefont {Lindsay}, \citenamefont
  {Broido}, \citenamefont {Gallego},\ and\ \citenamefont {Mingo}}]{mrl2016}%
  \BibitemOpen
  \bibfield  {author} {\bibinfo {author} {\bibfnamefont {J.}~\bibnamefont
  {Carrete}}, \bibinfo {author} {\bibfnamefont {W.}~\bibnamefont {Li}},
  \bibinfo {author} {\bibfnamefont {L.}~\bibnamefont {Lindsay}}, \bibinfo
  {author} {\bibfnamefont {D.~A.}\ \bibnamefont {Broido}}, \bibinfo {author}
  {\bibfnamefont {L.~J.}\ \bibnamefont {Gallego}}, \ and\ \bibinfo {author}
  {\bibfnamefont {N.}~\bibnamefont {Mingo}},\ }\href {\doibase
  10.1080/21663831.2016.1174163} {\bibfield  {journal} {\bibinfo  {journal}
  {Mater. Res. Lett.}\ }\textbf {\bibinfo {volume} {4}},\ \bibinfo {pages}
  {204} (\bibinfo {year} {2016})}\BibitemShut {NoStop}%
\bibitem [{\citenamefont {Kim}\ and\ \citenamefont
  {Majumdar}(2006)}]{kim_phonon_2006}%
  \BibitemOpen
  \bibfield  {author} {\bibinfo {author} {\bibfnamefont {W.}~\bibnamefont
  {Kim}}\ and\ \bibinfo {author} {\bibfnamefont {A.}~\bibnamefont {Majumdar}},\
  }\href {\doibase 10.1063/1.2188251} {\bibfield  {journal} {\bibinfo
  {journal} {J. Appl. Phys.}\ }\textbf {\bibinfo {volume} {99}},\ \bibinfo
  {pages} {084306} (\bibinfo {year} {2006})}\BibitemShut {NoStop}%
\bibitem [{\citenamefont {Genovese}\ \emph {et~al.}(2008)\citenamefont
  {Genovese}, \citenamefont {Neelov}, \citenamefont {Goedecker}, \citenamefont
  {Deutsch}, \citenamefont {Ghasemi}, \citenamefont {Willand}, \citenamefont
  {Caliste}, \citenamefont {Zilberberg}, \citenamefont {Rayson}, \citenamefont
  {Bergman},\ and\ \citenamefont {Schneider}}]{genovese2008}%
  \BibitemOpen
  \bibfield  {author} {\bibinfo {author} {\bibfnamefont {L.}~\bibnamefont
  {Genovese}}, \bibinfo {author} {\bibfnamefont {A.}~\bibnamefont {Neelov}},
  \bibinfo {author} {\bibfnamefont {S.}~\bibnamefont {Goedecker}}, \bibinfo
  {author} {\bibfnamefont {T.}~\bibnamefont {Deutsch}}, \bibinfo {author}
  {\bibfnamefont {S.~A.}\ \bibnamefont {Ghasemi}}, \bibinfo {author}
  {\bibfnamefont {A.}~\bibnamefont {Willand}}, \bibinfo {author} {\bibfnamefont
  {D.}~\bibnamefont {Caliste}}, \bibinfo {author} {\bibfnamefont
  {O.}~\bibnamefont {Zilberberg}}, \bibinfo {author} {\bibfnamefont
  {M.}~\bibnamefont {Rayson}}, \bibinfo {author} {\bibfnamefont
  {A.}~\bibnamefont {Bergman}}, \ and\ \bibinfo {author} {\bibfnamefont
  {R.}~\bibnamefont {Schneider}},\ }\href {\doibase 10.1063/1.2949547}
  {\bibfield  {journal} {\bibinfo  {journal} {The Journal of Chemical Physics}\
  }\textbf {\bibinfo {volume} {129}},\ \bibinfo {pages} {014109} (\bibinfo
  {year} {2008})}\BibitemShut {NoStop}%
\bibitem [{\citenamefont {Machado-Charry}\ \emph {et~al.}(2012)\citenamefont
  {Machado-Charry}, \citenamefont {Boulanger}, \citenamefont {Genovese},
  \citenamefont {Mousseau},\ and\ \citenamefont {Pochet}}]{machado2012}%
  \BibitemOpen
  \bibfield  {author} {\bibinfo {author} {\bibfnamefont {E.}~\bibnamefont
  {Machado-Charry}}, \bibinfo {author} {\bibfnamefont {P.}~\bibnamefont
  {Boulanger}}, \bibinfo {author} {\bibfnamefont {L.}~\bibnamefont {Genovese}},
  \bibinfo {author} {\bibfnamefont {N.}~\bibnamefont {Mousseau}}, \ and\
  \bibinfo {author} {\bibfnamefont {P.}~\bibnamefont {Pochet}},\ }\href
  {\doibase 10.1063/1.4754143} {\bibfield  {journal} {\bibinfo  {journal}
  {Appl. Phys. Lett.}\ }\textbf {\bibinfo {volume} {101}},\ \bibinfo {pages}
  {132405} (\bibinfo {year} {2012})}\BibitemShut {NoStop}%
\bibitem [{\citenamefont {Wang}\ \emph {et~al.}(2014)\citenamefont {Wang},
  \citenamefont {Pochet}, \citenamefont {Jenkins}, \citenamefont {Arenholz},
  \citenamefont {Bukalis}, \citenamefont {Gemming}, \citenamefont {Helm},\ and\
  \citenamefont {Zhou}}]{POC_PRB2014}%
  \BibitemOpen
  \bibfield  {author} {\bibinfo {author} {\bibfnamefont {Y.}~\bibnamefont
  {Wang}}, \bibinfo {author} {\bibfnamefont {P.}~\bibnamefont {Pochet}},
  \bibinfo {author} {\bibfnamefont {C.~A.}\ \bibnamefont {Jenkins}}, \bibinfo
  {author} {\bibfnamefont {E.}~\bibnamefont {Arenholz}}, \bibinfo {author}
  {\bibfnamefont {G.}~\bibnamefont {Bukalis}}, \bibinfo {author} {\bibfnamefont
  {S.}~\bibnamefont {Gemming}}, \bibinfo {author} {\bibfnamefont
  {M.}~\bibnamefont {Helm}}, \ and\ \bibinfo {author} {\bibfnamefont
  {S.}~\bibnamefont {Zhou}},\ }\href {\doibase 10.1103/physrevb.90.214435}
  {\bibfield  {journal} {\bibinfo  {journal} {Phys. Rev. B}\ }\textbf {\bibinfo
  {volume} {90}},\ \bibinfo {pages} {214435} (\bibinfo {year}
  {2014})}\BibitemShut {NoStop}%
\bibitem [{\citenamefont {Perdew}\ \emph {et~al.}(1996)\citenamefont {Perdew},
  \citenamefont {Burke},\ and\ \citenamefont {Ernzerhof}}]{vasp_pbe_1}%
  \BibitemOpen
  \bibfield  {author} {\bibinfo {author} {\bibfnamefont {J.~P.}\ \bibnamefont
  {Perdew}}, \bibinfo {author} {\bibfnamefont {K.}~\bibnamefont {Burke}}, \
  and\ \bibinfo {author} {\bibfnamefont {M.}~\bibnamefont {Ernzerhof}},\ }\href
  {\doibase 10.1103/PhysRevLett.77.3865} {\bibfield  {journal} {\bibinfo
  {journal} {Phys. Rev. Lett.}\ }\textbf {\bibinfo {volume} {77}},\ \bibinfo
  {pages} {3865} (\bibinfo {year} {1996})}\BibitemShut {NoStop}%
\bibitem [{\citenamefont {Willand}\ \emph {et~al.}(2013)\citenamefont
  {Willand}, \citenamefont {Kvashnin}, \citenamefont {Genovese}, \citenamefont
  {V{\'{a}}zquez-Mayagoitia}, \citenamefont {Deb}, \citenamefont {Sadeghi},
  \citenamefont {Deutsch},\ and\ \citenamefont {Goedecker}}]{newHGH2013}%
  \BibitemOpen
  \bibfield  {author} {\bibinfo {author} {\bibfnamefont {A.}~\bibnamefont
  {Willand}}, \bibinfo {author} {\bibfnamefont {Y.~O.}\ \bibnamefont
  {Kvashnin}}, \bibinfo {author} {\bibfnamefont {L.}~\bibnamefont {Genovese}},
  \bibinfo {author} {\bibfnamefont {{\'{A}}.}~\bibnamefont
  {V{\'{a}}zquez-Mayagoitia}}, \bibinfo {author} {\bibfnamefont {A.~K.}\
  \bibnamefont {Deb}}, \bibinfo {author} {\bibfnamefont {A.}~\bibnamefont
  {Sadeghi}}, \bibinfo {author} {\bibfnamefont {T.}~\bibnamefont {Deutsch}}, \
  and\ \bibinfo {author} {\bibfnamefont {S.}~\bibnamefont {Goedecker}},\ }\href
  {\doibase 10.1063/1.4793260} {\bibfield  {journal} {\bibinfo  {journal} {The
  Journal of Chemical Physics}\ }\textbf {\bibinfo {volume} {138}},\ \bibinfo
  {pages} {104109} (\bibinfo {year} {2013})}\BibitemShut {NoStop}%
\bibitem [{\citenamefont {Cresti}\ \emph {et~al.}(2003)\citenamefont {Cresti},
  \citenamefont {Farchioni}, \citenamefont {Grosso},\ and\ \citenamefont
  {Parravicini}}]{CRE_PRB68}%
  \BibitemOpen
  \bibfield  {author} {\bibinfo {author} {\bibfnamefont {A.}~\bibnamefont
  {Cresti}}, \bibinfo {author} {\bibfnamefont {R.}~\bibnamefont {Farchioni}},
  \bibinfo {author} {\bibfnamefont {G.}~\bibnamefont {Grosso}}, \ and\ \bibinfo
  {author} {\bibfnamefont {G.~P.}\ \bibnamefont {Parravicini}},\ }\href
  {\doibase 10.1103/PhysRevB.68.075306} {\bibfield  {journal} {\bibinfo
  {journal} {Phys. Rev. B}\ }\textbf {\bibinfo {volume} {68}},\ \bibinfo
  {pages} {075306} (\bibinfo {year} {2003})}\BibitemShut {NoStop}%
\bibitem [{\citenamefont {Togo}\ and\ \citenamefont {Tanaka}(2015)}]{phonopy}%
  \BibitemOpen
  \bibfield  {author} {\bibinfo {author} {\bibfnamefont {A.}~\bibnamefont
  {Togo}}\ and\ \bibinfo {author} {\bibfnamefont {I.}~\bibnamefont {Tanaka}},\
  }\href {\doibase 10.1016/j.scriptamat.2015.07.021} {\bibfield  {journal}
  {\bibinfo  {journal} {Scr. Mater.}\ }\textbf {\bibinfo {volume} {108}},\
  \bibinfo {pages} {1} (\bibinfo {year} {2015})}\BibitemShut {NoStop}%
\bibitem [{\citenamefont {Kresse}\ and\ \citenamefont
  {Furthm\"uller}(1996)}]{vasp_general_4}%
  \BibitemOpen
  \bibfield  {author} {\bibinfo {author} {\bibfnamefont {G.}~\bibnamefont
  {Kresse}}\ and\ \bibinfo {author} {\bibfnamefont {J.}~\bibnamefont
  {Furthm\"uller}},\ }\href {\doibase 10.1103/PhysRevB.54.11169} {\bibfield
  {journal} {\bibinfo  {journal} {Phys. Rev. B}\ }\textbf {\bibinfo {volume}
  {54}},\ \bibinfo {pages} {11169} (\bibinfo {year} {1996})}\BibitemShut
  {NoStop}%
\bibitem [{\citenamefont {Bl\"ochl}(1994)}]{vasp_paw_1}%
  \BibitemOpen
  \bibfield  {author} {\bibinfo {author} {\bibfnamefont {P.~E.}\ \bibnamefont
  {Bl\"ochl}},\ }\href {\doibase 10.1103/PhysRevB.50.17953} {\bibfield
  {journal} {\bibinfo  {journal} {Phys. Rev. B}\ }\textbf {\bibinfo {volume}
  {50}},\ \bibinfo {pages} {17953} (\bibinfo {year} {1994})}\BibitemShut
  {NoStop}%
\bibitem [{\citenamefont {Kresse}\ and\ \citenamefont
  {Joubert}(1999)}]{vasp_paw_2}%
  \BibitemOpen
  \bibfield  {author} {\bibinfo {author} {\bibfnamefont {G.}~\bibnamefont
  {Kresse}}\ and\ \bibinfo {author} {\bibfnamefont {D.}~\bibnamefont
  {Joubert}},\ }\href {\doibase 10.1103/PhysRevB.59.1758} {\bibfield  {journal}
  {\bibinfo  {journal} {Phys. Rev. B}\ }\textbf {\bibinfo {volume} {59}},\
  \bibinfo {pages} {1758} (\bibinfo {year} {1999})}\BibitemShut {NoStop}%
\bibitem [{\citenamefont {Katre}\ \emph {et~al.}(2017)\citenamefont {Katre},
  \citenamefont {Carrete}, \citenamefont {Dongre}, \citenamefont {Madsen},\
  and\ \citenamefont {Mingo}}]{Katre_PRL17}%
  \BibitemOpen
  \bibfield  {author} {\bibinfo {author} {\bibfnamefont {A.}~\bibnamefont
  {Katre}}, \bibinfo {author} {\bibfnamefont {J.}~\bibnamefont {Carrete}},
  \bibinfo {author} {\bibfnamefont {B.}~\bibnamefont {Dongre}}, \bibinfo
  {author} {\bibfnamefont {G.~K.~H.}\ \bibnamefont {Madsen}}, \ and\ \bibinfo
  {author} {\bibfnamefont {N.}~\bibnamefont {Mingo}},\ }\href {\doibase
  10.1103/PhysRevLett.119.075902} {\bibfield  {journal} {\bibinfo  {journal}
  {Phys. Rev. Lett.}\ }\textbf {\bibinfo {volume} {119}},\ \bibinfo {pages}
  {075902} (\bibinfo {year} {2017})}\BibitemShut {NoStop}%
\end{thebibliography}
\end{document}